\documentclass[a4paper,fleqn,usenatbib,useAMS]{mnras}
\usepackage[final]{graphicx}
\usepackage{amssymb}
\usepackage{amsmath}
\usepackage{lastpage}
\usepackage{dsfont}
\usepackage{float}
\usepackage{tabularx}
\usepackage{pdfpages}
\usepackage{color}
\usepackage[toc,page]{appendix}
\usepackage{gensymb}
\usepackage{epstopdf, epsfig}
\usepackage{hyperref}
\usepackage{pgfplots}
\pgfplotsset{compat=newest}
\usepackage{multicol}  
\usepackage{bm}	
\usepackage{pdflscape}
\usepackage[T1]{fontenc}
\usepackage{ae,aecompl}

\title{Stellar encounters with Giant Molecular Clouds}
\date{\today}
\author[G. Kokaia, M. B. Davies]{Giorgi Kokaia$^{1}$\thanks{Contact e-mail: \href{mailto:giorgi@astro.lu.se}{giorgi@astro.lu.se}}, Melvyn B. Davies$^{1}$
\\
$^{1}$Lund Observatory, Department of Astronomy and Theoretical Physics, Lund University, Box 43, SE-221 00 Lund, Sweden}
\date{\today}

\pubyear{2018}

\begin{document}
\label{firstpage}
\pagerange{\pageref{firstpage}--\pageref{lastpage}}
\maketitle

\begin{abstract}
Giant molecular clouds (GMCs) are believed to affect the biospheres of planets as their host star passes through them. We simulate the trajectories of stars and GMCs in the Galaxy and determine how often stars pass through GMCs. We find a strong decreasing dependence with Galactocentric radius, and with the velocity perpendicular to the Galactic plane, $V_z$.  The \textit{XY}-component of the kinematic heating of stars was shown to not affect the GMC hit rate, unlike the \textit{Z}-dependence ($V_z$) implies that stars hit fewer GMCs as they age. GMCs are locations of star formation, therefore we also determine how often stars pass near supernovae. For the supernovae the decrease with $V_z$ is steeper as how fast the star passes through the GMC determines the probability of a supernova encounter. We then integrate a set of Sun-like trajectories to see the implications for the Sun. We find that the Sun hits $1.6\pm1.3$ GMCs per Gyr which results in $1.5\pm1.1$ or (with correction for clustering) $0.8\pm 0.6$ supernova closer than 10 pc per Gyr. The different the supernova frequencies are from whether one considers multiple supernova per GMC crossing (few Myr) as separate events.  We then discuss the effect of the GMC hits on the Oort cloud, and the Earth's climate due to accretion, we also discuss the records of distant supernova. Finally, we determine Galactic Habitable Zone using our model. For the thin disk we find it to lie between 5.8-8.7 kpc and for the thick disk to lie between 4.5-7.7 kpc. 
\end{abstract}

\section{Introduction}
As stars orbit in the Galaxy they interact with giant molecular clouds (GMC). GMCs are large and massive gas clouds with masses ranging from tens of thousands to several million solar masses with sizes between 10 and 100 pc and there are upwards of 8000 of them in the Milky Way \citep{Miville-Deschenes2017}; they are primarily located inside the spiral arms, because that is where the gas can cool to form them \citep{Hou2009}. Because there are so many of them and because they are so large and massive  it is inevitable that stars will interact with them. This interaction can imply two things. If the star passes the GMC at a distance it will be scattered, contributing to the kinematic heating of a star as it ages \citep{Nordstrom2004}. If it gets so close that it passes through the GMC it will not only be scattered as a whole range of other things can occur during the passage.

Passing through the GMC can potentially be catastrophic for the habitability of a planet orbiting the star as it represents a large deviation from normal environmental conditions. This can lead to a so called mass extinction (large loss of bio-diversity) for the following reasons. The GMCs are massive and exert a strong tidal force on a structure such as the Oort Cloud \citep{Oort1950}, believed to exist around most other stars \citep{Tremaine1993}. This interaction has been shown to be strong enough to unbind comets orbiting in it \citep{Napier1982, Hut1985}. These comets could be injected into the planetary system leading to giant impacts \citep{Rubincam2016}. Or, as the star passes through the GMC upwards of an Earth mass of gas can be accreted \citep{Hoyle1939}. This gas can settle in the upper atmosphere of planets and when doing so can seed rapid cloud formation \citep{Kataoka2013} or enshroud the upper atmosphere \citep{Pavlov2005, Nimura2015}, either of these will cause a sharp increase in albedo and can lead to a so called ``Snowball Earth" scenario.

An additional threat is posed by the GMC passage due to them being sites of star formation. Stars that end their lives as core-collapse supernovae are short-lived ($<50$ Myrs, see for example \citealp{Zapartas2017}) and therefore end their lives and explode where or close to where they are born. This means that whilst passing through a GMC a star risks passing near a supernova. There are several ways in which a nearby supernova can be catastrophic for life on the surface of a planet. (1) Assuming the planet has an ozone layer protecting it from high energy radiation, \cite{Beech2011} shows that a typical $10^{46}$ J supernova (giving out $\sim10^{38}$ W of FUV and X-rays during the initial shock break out for a few minutes) within $9$ pc can cause mass extinction due to UV and X-ray irradiation of the surface even with the ozone layer absorbing 99\% of the radiation. Barring such protection, he places the ``kill radius" at 90 pc. (2) \cite{Ruderman1974} discusses a mechanism in which this protective layer can be destroyed by the formation of nitric oxide (NO) in the upper atmosphere which then will destroy the ozone and cause the UV flux of the host star to cause the mass extinction. Taking the Sun as an example, it has an FUV/X-ray power output of $10^{20}$ W \citep{Gudel2007}, meaning the irradiation of the Earth will be a factor $10^3$ lower than that of the \cite{Beech2011} supernova, however the ozone will take decades to replenish \citep{Thomas2005} meaning the overall irradiation will be significantly higher as the relative duration differs by a factor of $10^6$. For the destruction of the ozone layer to occur, \cite{Ruderman1974} assumes a supernova producing $10^{40}$ J of X-rays or $10^{43}-10^{44}$ J of cosmic rays needs to occur at a distance of $\sim$ 15 pc. (3) Further, \cite{Thomas2005} showed (in the context of a 2 kpc distant GRB, producing equivalent amounts of FUV and X-rays hitting the Sun as a nearby supernova) how the production of NO can lead to an extinction from an ice age rather than irradiation. The NO will form nitrogen dioxide (NO${}_2$) which will significantly increase the albedo of the planet plunging it into an ice age. Not only that, but the biosphere would also be stressed by nitric acid rain. (4) Similarly, \cite{Tanaka2006} showed that at a distance of 12-15 pc $10^{43}$ J of cosmic rays can also cause an ice age by seeding cloud formation in the upper atmosphere. Which one of these effects will cause the mass extinction in the event of a nearby supernova is up for debate but it will likely depend on the distance to the supernova and the conditions in which it occurs. These works consider all consider the Sun completely unshielded, however the nearby supernova we will discuss in this paper occur within GMCs. The amount of shielding will depend strongly on what kind of GMC the supernova occurs in and where in the GMC it happens. It is clear that there will be some shielding regardless and for that reason we adopt a 10 pc kill radius.
 
In this work we will focus on the destructive effects of a star passing through a GMC or being near a supernova as it detonates. It is important to point out that there is a different way in which one can study the effect of the Galactic environment on the habitability of planetary systems, in particular the effects of cosmic rays. Instead of looking at the cosmic rays originating from a nearby supernova one can look at the integrated flux from a larger number of more distant supernovae and their remnants. \cite{Shaviv2002, Shaviv2003} does this by constructing a diffusion model of the Galaxy in which the cosmic ray sources are the spiral arms since that is where nearly all the supernova are found. They find that the cosmic ray flux varies by up to a factor of four between the minima in the inter-arm region and the maxima when crossing the spiral arm and suggest that this variation could be the cause the historical glaciations observed with some regularity in the geological archive.

The question of how often the Sun passes through a GMC or encounters a nearby supernova was thrust into the limelight when \cite{Raup1984} saw a periodicity in mass extinctions in the geological record and speculated it could have an extra-Solar origin. For an in-depth review on this topic in particular see \cite{Bailer-Jones2009}. More generally, it ties into the discussions of the Galactic Habitable Zone. The Galactic Habitable Zone (GHZ) is defined as the annular region in which the Galaxy is most habitable and likely to host life. The outer boundary is set by looking at the average metallicity of stars. \cite{Gonzalez2001} estimates that a metallicity of [Fe/H]$\gtrsim -0.7$ is needed to form Earth-like planets, i.e. planets with mechanisms for interior heat loss, volatile inventory for complex life and sufficient atmospheric retention. The inner boundary is usually set in one of two ways: The first way is to set the boundary where the metallicity gets too high \citep{Gonzalez2001, Vuktoic2018}, because at that point you will also start producing a lot of gas giants \citep{Fischer2005} instead of just rocky planets and the rocky planets you produce will have too high surface gravities to be habitable by organic life. It is known from e.g. \cite{Carrera2016} that a high multiplicity of gas giants cause problems for habitable systems as they destabilize the system. The other way of determining the boundary is to look at supernova rates \citep{Gowanlock2011, Spitoni2014, Morrison2015}, and then calculate how often a star would experience a nearby supernova. One then sets a threshold for the acceptable supernova rate (usually a factor of the nearby Solar supernova rate) and find the boundary. Since both the average metallicty and supernova rate in a region of the Galaxy will change with time one can actually find a time-dependant GHZ such as \cite{Lineweaver2004}.  

In this paper we will determine how often stars pass through GMCs by integrating their orbits and counting how often they coincide. In section \ref{sec:two} the Galactic potential used for integrating orbits of stars and GMCs will be discussed. We also discuss the spiral arms of the Galaxy and how they're used in the simulations along with how we use Galactic observations of ${\rm H}_2$ to synthesize the population of GMCs. Following that we give the results in \ref{sec:res}. In section \ref{sec:four} we add supernovae into the GMCs, discuss the implications for the Sun and finish off by determining the GHZ using what we find in this paper together with the methods discussed in papers mentioned above.

\section{Setup of the Numerical Experiment}\label{sec:two}
\subsection{Galactic Potential}
We conduct an experiment in which we follow orbits of stars and GMCs to investigate how often stars on different orbits pass through GMCs, which we from now on will refer to as "hitting" the GMCs. We do this by integrating them through a Galactic potential. Following \cite{Feng2013} we use a three-component symmetric, analytic Galactic potential given in \cite{Juarez2010}. The potential $\Phi_G(R, z)$ is given by

\begin{equation}
\Phi_G = \Phi_{\rm b}+\Phi_{\rm h}+\Phi_{\rm d},
\end{equation}
\noindent The three components are the halo, bulge and disk. The halo and bulge are modelled as two spherical Plummer \citep{Plummer1911} potentials. The equation describing them is shown below in cylindrical coordinates.
\begin{equation}
\Phi_{\rm b,h} = -\frac{GM_{\rm b,h}}{\sqrt{R^2+z^2+b^2_{\rm b,h}}},
\label{eq:plum}
\end{equation}
\noindent The disk we model as a  Miyamoto \citep{Miyamoto1975} disk, shown below.
\begin{equation}
\Phi_{\rm d} = -\frac{GM_{\rm d}}{\sqrt{R^2+\left(a_{\rm d}+\sqrt{z^2+b_{\rm d}^2}\right)^2}}
\label{eq:disk}
\end{equation}
\noindent The above equations are in cylindrical coordinates, meaning $R$ represents the Galactocentric distance and $z$ the distance from the Galactic plane. $M_{\rm d, b, h}$ is the mass of the disk, bulge and halo respectively. $b$ represents the scale length of each component and $a_d$ is the scale height of the disk. The values for these parameters can be found in table \ref{tab:Potparam}.\\
\begin{table}
\centering
\caption{The parameter values used in equations \protect\ref{eq:plum} and \protect\ref{eq:disk} to construct the potential which is used in the integration \protect\cite{Juarez2010}.}
\begin{tabularx}{\columnwidth}{l  l}
\hline
\small Parameter & Value\\
\hline
\small $M_{\rm d}$ & $7.9080\times10^{10}~M_\odot$ \\
\small $M_{\rm b}$ & $1.3955\times10^{10}~M_\odot$ \\
\small $M_{\rm h}$ & $6.9766\times10^{11}~M_\odot$ \\
\small $a_{\rm d}$ & 3550 pc \\
\small $b_{\rm b}$ & 250 pc \\
\small $b_{\rm d}$ & 350 pc \\
\small $b_{\rm h}$ & 24000 pc \\
\hline
\end{tabularx}
\label{tab:Potparam}
\end{table}

\noindent Some tests were also conducted using other potentials, such as \cite{Paczynski1990} and \cite{Millan2017}. We found that the number of GMC hits changed by a factor of $\sim$10\% for orbits with the same set of initial conditions in the different potentials, overall however the number of hits remain unchanged. This due to the fact that changing potential results in changing the orbit of both stars and GMCs in the same way. More on this in section \ref{sec:pots}.

\subsection{Spiral arms}\label{sec:sa}
\begin{figure}
\centering
\includegraphics[width = 1\columnwidth]{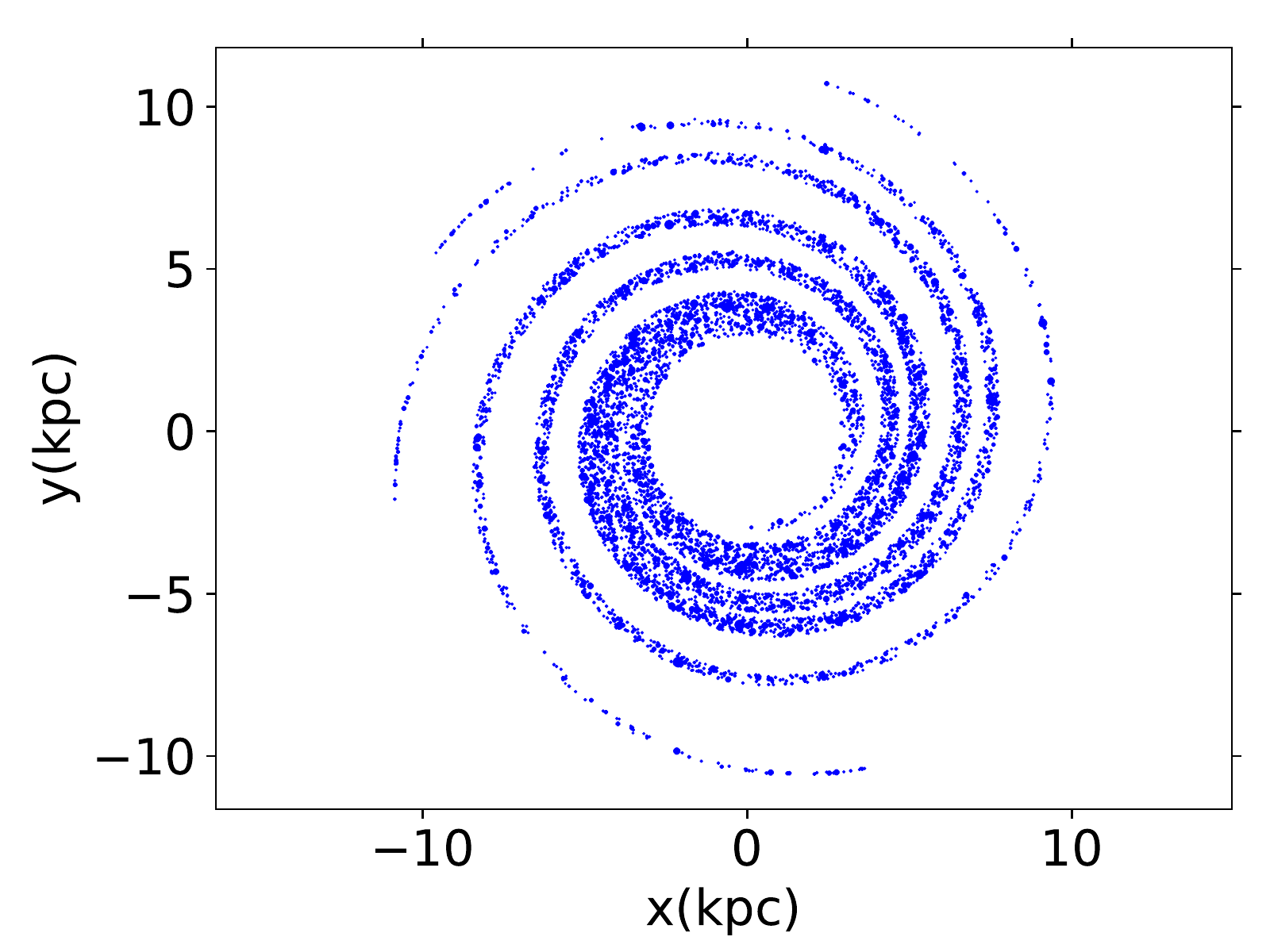}
\caption{Distribution of GMCs in the Galaxy. The arms are generated using equation \protect\ref{eq:arm} and the values in table \protect\ref{tab:Armparam}. The arms range from 3 to 12 kpc with  a pattern speed of $20~{\rm km/s~kpc}^{-1}$, measured pattern speeds are given in table \protect\ref{tab:patterns}. Note that GMC sizes in figure are to scale. }
\label{fig:gmcdist}
\end{figure}

In the Galaxy, GMCs are primarily located in the spiral arms. In fact, they are one of the best ways to find and trace the arms \citep{Hou2009, Hou2014}. We do not consider the spiral arm gravitational potential, however to accurately reproduce the Galactic GMC population in our experiment we distribute them in the spiral pattern observed by \cite{Hou2014}. These observations are carried out in three different ways: 1) They look for HII regions, because GMCs are regions of star formation and young, hot stars will ionize the surrounding gas, 2) The study includes observations of methanol masers which also trace star formation, 3) Finally, they look for CO emission. CO is a very fragile molecule that cannot survive unless embedded in another, dense gas. It is therefore an excellent tracer for GMCs. To the observations they then fit a logarithmic spiral, shown below
\begin{equation}
\ln\frac{R}{R_{\rm i}}=(\theta-\theta_{\rm i})\tan\psi_{\rm i}
\label{eq:arm}
\end{equation}
Where $R_{\rm i}$ is the starting radius, $\theta_{\rm i}$ the starting angle and $\psi_{\rm i}$ the pitch angle, i.e. how non-circular the spiral is. $\theta$ and $R$ are cylindrical coordinates. The best fit to their observations is a four-armed spiral and the parameters can be found in table \ref{tab:Armparam}. We use these parameters for all our models (as seen in table \ref{tab:models}) except one which uses a three arm fit from the same paper. Parameters for the three armed spiral are shown in table \ref{tab:3param}. The GMC distribution generated from table \ref{tab:Armparam} can be seen in figure~\ref{fig:gmcdist}.

The spiral pattern rotates rigidly in the Galaxy, i.e. at a constant angular velocity. This angular velocity is known as the pattern speed. Some measured values for this pattern speed are shown in table \ref{tab:patterns}. This motion around the Galaxy is very different from that of stars that have a fairly flat rotation curve due to the Galactic potential. A flat rotation curve implies a constant circular velocity which means that the angular velocity will be linearly decreasing, so stars at different radii will have a different velocity relative to the spiral arms. The radius at which the angular velocities coincide is known as the corotation radius.
\begin{table}
\centering
\caption{Arm parameters used in equation \protect\ref{eq:arm} in order to construct the spiral arms used in the simulation models with four arms. The Rs determine at which radius each arm starts, the $\theta$ determines at what angle they start and $\psi$ is the pitch angle, i.e. how non-circular each arm is. Values are from \protect\cite{Hou2014}.}
\begin{tabularx}{\columnwidth}{l l l l l }
\hline
  & \textbf{Arm 1} & \textbf{Arm 2} & \textbf{Arm 3} &  \textbf{Arm 4}\\
\hline
 $R$ & $3270$ pc & $4290$ pc & $3580$ pc  &  $3980$ pc  \\
 $\theta$ & $38.5^\circ$  & $189.0^\circ$ & $215.2^\circ$ & $320.0^\circ$ \\
 $\psi$ & $9.87^\circ$ & $10.51^\circ$   & $10.01^\circ$   & $8.14^\circ$ 
\end{tabularx}
\label{tab:Armparam}
\end{table}
\begin{table}
\centering
\caption{Measurements of the pattern speed of the spiral arms in the Milky Way.}

\begin{tabularx}{\columnwidth}{l l}
\hline
\small Study & Result\\
\hline
\small \cite{Martos2004} &  $\Omega=20~\mbox{km s}^{-1}~\mbox{kpc}^{-1}$ \\
\small \cite{Bissantz2017}  & $\Omega=20~\mbox{km s}^{-1}~\mbox{kpc}^{-1}$ \\
\small \cite{Li2016}  & $\Omega=23\pm2~\mbox{km s}^{-1}~\mbox{kpc}^{-1}$ \\
\small \cite{Junqueira2017} & $\Omega=23\pm1~\mbox{km s}^{-1}~\mbox{kpc}^{-1}$ \\
\small \cite{Dias2005} & $\Omega=24~\mbox{km s}^{-1}~\mbox{kpc}^{-1}$\\
\small \cite{Vallee2017} & $\Omega=23\pm2~\mbox{km s}^{-1}~\mbox{kpc}^{-1}$\\
\hline
\end{tabularx}
\label{tab:patterns}
\end{table}

\subsection{GMC properties}\label{sec:gmc}
To model the GMC population in addition to the spiral distribution of the GMCs we also need their remaining spatial distribution $(R, z)$, the distribution of their sizes and densities, their mass function and their lifetimes. For the mass function we use equation \ref{eq:dndm} fit to observations by \cite{Rosolowsky2006} as shown below.
\begin{equation}
\frac{dN}{dM}=(\gamma+1)\frac{N_0}{M_0}\left(\frac{M}{M_0}\right)^\gamma,~~M<M_0
\label{eq:dndm}
\end{equation}
\noindent Where $\gamma=1.53$ and $N_0=36$ are constants. $M_0$ is the maximum mass of $3\times10^6{\rm M}_\odot$. The minimum mass is $10^4{\rm M}_\odot$. The resulting mass distribution ends up being fairly top heavy, with 3.5\% of GMCs above $10^6{\rm M}_\odot$, accounting for 40\% of the total ${\rm H}_2$ mass. 

A GMC forms slowly as gas falls into the local spiral arm potential and accumulates. It then starts losing mass due to stellar feedback from the newly formed stars as it is leaving the spiral arm potential. We model this behaviour by adopting an age-dependant mass, following the prescription in \cite{Gustafsson2016}. GMC mass varies with age as shown below
\begin{equation}
M(t) = \left(-0.25\times\left(\frac{t-t_0}{10}\right)^2+\frac{t-t_0}{10}\right)\times M_i
\label{eq:mt}
\end{equation}
\noindent where $M_i$ is the mass drawn from equation \ref{eq:dndm}, $t-t_0$ is the age at time t in Myrs. This results in a 40 Myr lifetime where the GMC increases in mass for the first 20 Myrs, peaks and then decreases for another 20 Myrs. The 40 Myr lifetime is consistent with observations from e.g. \cite{Williams1997}, \cite{Ferrara2016}, and the shape of the function given by equation \ref{eq:mt} is based on simulations from \cite{Krumholz2006} and \cite{Goldbaum2011}.

\begin{figure}
\centering
\includegraphics[width = 1\columnwidth]{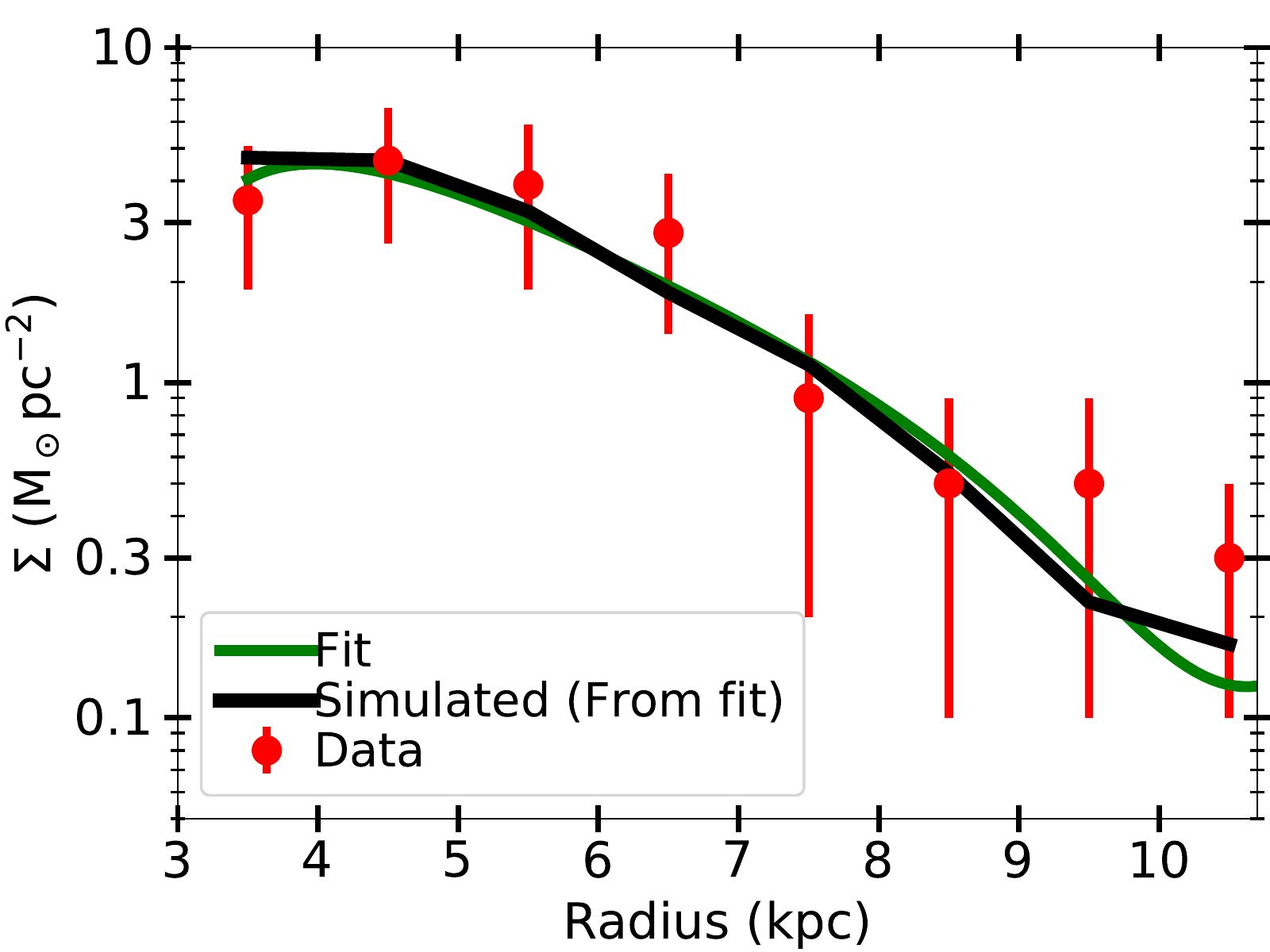}
\caption{GMC surface density as a function of Galactocentric radius. The black line shows the average GMC surface density measured in our simulations. The distribution of the clouds were generated using the data of \protect\cite{Nakanishi2006}, which is shown as the red points in the figure (values listed in table \protect\ref{tab:naka}). The green line shows the fitted function from which the data is generated, note that the black and green line diverge at large R, this is due to the small number count of clouds there. Results are also consistent with other work such as \protect\cite{Clemens1988},  \protect\cite{Scoville1987},  \protect\cite{Wienen2015},  \protect\cite{Roman-Duval2016}.}
\label{fig:surf}
\end{figure}

Given a mass drawn using equation \ref{eq:dndm} and adjusted according to equation \ref{eq:mt} we determine the GMC radius by first putting the modelled GMCs in one of three populations. Following the categorization and nomenclature from observations by \cite{Roman-Duval2016} the ${\rm H}_2$ gas is put in one of three categories; very dense, dense and diffuse. They make up 15, 60 and 25\% of all the gas, respectively. For the very dense gas we already have a mass-radius relationship from \cite{Roman-Duval2010} shown in equation \ref{eq:mr}, this gives the range of gas surface densities between 130 and 300 ${\rm M}_\odot {\rm pc}^{-2}$.
\begin{equation}
\left(\frac{R_{\rm GMC}}{1~{\rm pc}}\right) = \frac{1}{229}\left(\frac{M}{1~M_\odot}\right)^{1/2.36}
\label{eq:mr}
\end{equation}
\noindent We then construct two additional mass-radius relationships to get full range of observed surface densities in \cite{Roman-Duval2016} and make sure it is consistent with other observations such as \cite{Miville-Deschenes2017}. The two additional mass-radius relationships are shown below. Equation \ref{eq:mr2} covers the range from 50 to 130 ${\rm M}_\odot {\rm pc}^{-2}$, equation \ref{eq:mr3} covers 25 to 50 ${\rm M}_\odot {\rm pc}^{-2}$.
\begin{equation}
\left(\frac{R_{\rm GMC}}{1~{\rm pc}}\right) = \frac{1}{79}\left(\frac{M}{1~M_\odot}\right)^{1/2.36}
\label{eq:mr2}
\end{equation}
\begin{equation}
\left(\frac{R_{\rm GMC}}{1~{\rm pc}}\right)= \frac{1}{28}\left(\frac{M}{1~M_\odot}\right)^{1/2.36}
\label{eq:mr3}
\end{equation}
The radial and vertical distributions are given by ${\rm H}_2$ observations by \cite{Nakanishi2006}. Their observed data is shown in table \ref{tab:naka}. We take the radial data and to it we fit a 6th degree polynomial using the \texttt{numpy} module \texttt{polyfit}, using this we then generate our own radial distribution with an accept/reject-method. The resulting (time averaged) surface distribution can be seen in figure~\ref{fig:surf} along with the observed data and the fitted polynomial. The fitted constants to the polynomial, shown in equation~\ref{eq:poly}, are given in table \ref{tab:poly}.
\begin{equation}
y = \sum_{n=0}^6 k_n x^n
\label{eq:poly}
\end{equation}


\begin{table}
\centering
\caption{Measurements of molecular gas in the Galaxy, from \protect\cite{Nakanishi2006}. Measurements are averages over binned annuli and $\sigma$ is the standard deviation of a Gaussian distribution fitted to the gas density about the midplane.}
\begin{tabularx}{\columnwidth}{ l  l  l}
\hline
\small Radius (kpc) & Surface density ($M_\odot{\rm pc}^{-2}$) & $\sigma$ (pc)\\
\hline
3-4 & 3.5$\pm$1.6 & 37$\pm$10\\
4-5 & 4.6$\pm$2.0 & 33$\pm$5\\
5-6 & 3.9$\pm$2.0 & 43$\pm$4\\
6-7 & 2.8$\pm$1.4 & 37$\pm$9\\
7-8 & 0.9$\pm$0.7 & 38$\pm$13\\
8-9 & 0.5$\pm$0.4 & 79$\pm$8\\
9-10 & 0.5$\pm$0.4 & 78$\pm$8\\
10-11 & 0.3$\pm$0.2 & 69$\pm$8\\
\hline
\end{tabularx}

\label{tab:naka}

\end{table}
\begin{table}
\centering
\caption{The factors in the 6th degree polynomial plotted using equation~\ref{eq:poly} to generate the surface density (${\rm M}_\odot {\rm pc}^{-2}$) as a function of radius (kpc) in figure~\ref{fig:surf}.}
\begin{tabularx}{\columnwidth}{l r r r r r r r r}
\hline
\small $n$ & 0 & 1 & 2 & 3 & 4 & 5 & 6\\
\small $k_n$ & -106 & 91 & -30 & 5.2 & -0.50 & 0.0025 & -5.3$\times10^{-4}$\\
\hline
\end{tabularx}
\label{tab:poly}
\end{table}
\subsection{Integration of orbits within the Galaxy}\label{sec:Pot}
We populate our simulation with 6 700 randomly generated clouds and keep this number constant; as GMCs reach the end of their life a new one is initiated in the next time-step. At the start of the simulation the clouds are given a random age and their mass is changed according to equation \ref{eq:mt}, this results in a total ${\rm H}_2$ mass of $\sim6.5\times10^8{\rm M}_\odot$ in agreement with the total ${\rm H}_2$ mass as measured with CO observations by \cite{Roman-Duval2016}. The initial conditions are set as follows: the Galactocentric radius, $R$, is chosen by using the accept/reject method with the function in figure~\ref{fig:surf}, $z$ and $V_z$ chosen such that observed scale height from \cite{Nakanishi2006} is maintained (values from observations are shown in table \ref{tab:naka})

$\theta$ is chosen such that the cloud crosses the analytic line of a randomly chosen spiral arm 20 Myrs into its 40 Myr lifetime, i.e. when it is at peak mass. The velocities are chosen such that the GMCs end up on circular orbits. The resulting distribution of GMCs, using Model A listed in table \ref{tab:models} can be seen in figure~\ref{fig:gmcdist}. The way we model the GMC trajectories and lifetimes (equation~\ref{eq:mt}) results in the width of the spiral arms being a function of the relative difference between the local orbital velocity and the pattern speed of the spiral arms. The difference in width between the inner and outer arms can be seen in figure~\ref{fig:gmcdist}. In it the arms are a few kpc wide in the inner region and a few 100 pc in the outer regions.

We also perform the experiments with different sets of spiral arms structures, primarily by varying the pattern speed; the different models can be seen in table \ref{tab:models}. The parameters for the three armed spiral in model D is shown in table \ref{tab:3param}. Model E was to test how the experiment scales with increased total GMC mass. This also implies a higher star formation rate, which we know the Galaxy had in the past from e.g. \cite{vanDokkum2013} who show that as a function of redshift the star formation rate changes as $log(1+{\rm SFR}) = 0.26 + 0.92z - 0.23z^2$, where the redshift at the birth of the Sun (4.5 Gyrs ago) is 0.415. 

For the stellar trajectories, we initialize 50 000 runs each at 8 different radii, $R=(5, 5.5, 6, 6.5, 7, 7.5, 8, 8.5)$kpc. The polar angle, $\theta$, was randomly, uniformly chosen between 0 and 2$\pi$, we do 1000 runs for each integer $V_z$ between 1 and 50 km/s where $V_z$ is the velocity away from the Galactic plane defined at $z=0$. Unlike the GMCs the stars are not on circular orbits. We tested setting the initial velocities with different velocity dispersions $\sigma_\theta$ and $\sigma_R$. These are standard deviations in the normal distribution from which $V_R$ and $V_{\rm \theta}$ are randomly generated. These velocities can be characterized as the non-circular component of the orbital velocity, for $V_{\rm \theta}$ we then have; $V_{\rm orb} = V_{\rm circ}+V_\theta$. The sets of velocity dispersions used were: $\sigma_{\rm R}=[20.0, 22.5, 25.0, 27.5, 30.0]$ and $\sigma_{\rm \theta} = [10.0, 12.5, 15.0, 17.5, 20]$. In most runs we used $\sigma_{\rm R} = 25$ and $\sigma_{\rm \theta} = 15$ km/s. Corresponding to a 4 Gyr old star in the Solar neighbourhood.

For the integration we use a leapfrog integrator with variable timestep, where the maximum timestep is $h=0.1$ Myr. To make sure the scattering during close encounters are well modelled, when a star gets close to a GMC it is changed as follows:
\begin{equation}
h = \frac{0.1\left({\rm int}(d/R_{\rm GMC})+1\right)}{10}~{\rm Myr}
\end{equation}
\noindent where d is separation between star and GMC. We also make sure that the relative velocity of the star to the GMC is always larger than the escape velocity. This because the changing mass and the fact that GMCs can form very close to stars could lead to unphysical encounters. We find that these are very rare and for the few that were found we removed the trajectory for the analysis. 

The duration of the integration is 1 Gyr. There are three reasons for choosing 1 Gyr as the integration time. First, it is sufficiently long to complete one revolution around the Galaxy at the outermost radius in the spiral arm rest frame. Second, it is the time-scale for which the static, unchanging Galaxy and the spiral arm model remains valid \citep{Sellwood2014}. Finally, since we set initial conditions for a range of $V_z$ this can be viewed as sampling a Gyr snapshot of stars of different ages and different interaction histories with GMCs. This because the GMC interaction will primarily act to drive up the $V_z$ \citep{Sellwood2002}. 
\begin{table}
\centering
\caption{The different sets of parameters for the GMC distributions in the simulation. The pattern speeds in model B and C are calculated and set up such that the co-rotation radii are at 8 and 6 kpc respectively and in the others we see it is consistent with values from table \ref{tab:patterns} shows a list of measured pattern speeds}
\begin{tabularx}{\columnwidth}{l l l}
\hline
\small Model & Arm properties & Pattern Speed ($~\mbox{km s}^{-1}\mbox{kpc}^{-1}$)\\
\hline
\small A  & 4-arm spiral & 20$\rightarrow$co-rotation: 11.6 kpc\\
\small B  & 4-arm spiral  & 28.8$\rightarrow$co-rotation: 8.0 kpc\\
\small C  & 4-arm spiral  & 38.6$\rightarrow$co-rotation: 6.0 kpc\\
\small D  & 3-arm spiral & 20$\rightarrow$co-rotation: 11.6 kpc\\
\small E  & 4-arm spiral with \\ & double GMC mass & 20$\rightarrow$co-rotation: 11.6 kpc\\
\small F  & No arms & -\\
\hline
\end{tabularx}

\label{tab:models}
\end{table}
\begin{table}
\centering
\caption{Arm parameters used in equation \protect\ref{eq:arm} with an alternate, three-armed spiral mode used in model D. This is an alternate fit to the data, presented in \protect\cite{Hou2014}. It gives slightly worse fit, but it's interesting to see if any differences are observed using three arms instead of four.}
\begin{tabularx}{\columnwidth}{l l l l }
\hline
  & \textbf{Arm 1} & \textbf{Arm 2} & \textbf{Arm 3}\\
\hline
 $R$ & $3220$ pc & $3430$ pc & $3100$ pc    \\
 $\theta$ & $44.1^\circ$  & $184.5^\circ$ & $210.3^\circ$ \\
 $\psi$ & $9.25^\circ$ & $9.50^\circ$   & $7.80^\circ$   
\end{tabularx}
\label{tab:3param}
\end{table}
\section{Results}\label{sec:res}
\begin{figure}
\centering
\includegraphics[width = 1\columnwidth]{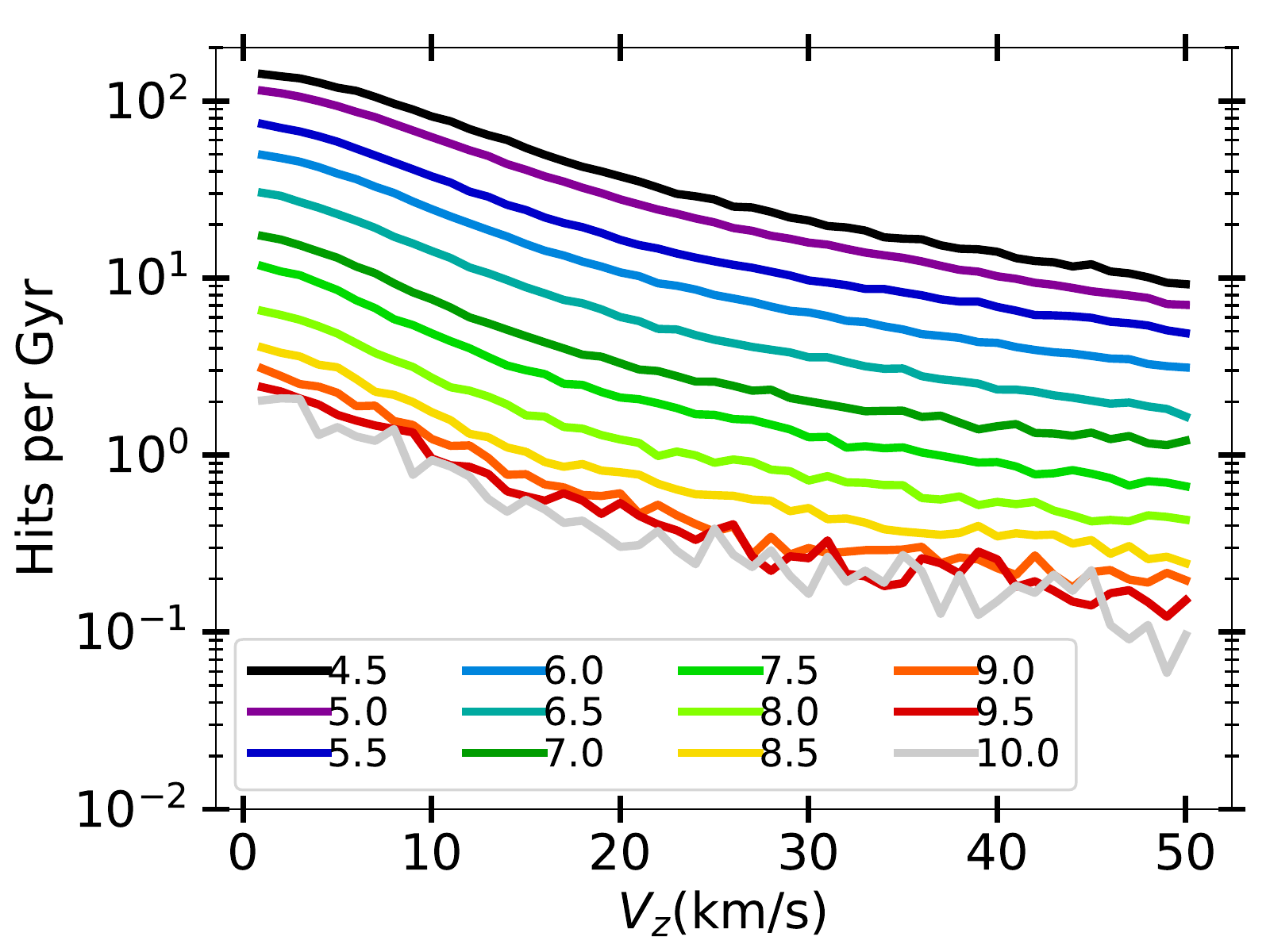}
\caption{The mean number of GMC hits per Gyr as a function of initial $V_z$. The stars were initialized at different radii using spiral arm model A and then binned and plotted in 500 pc wide bins centred at the values shown in the figure. Velocity dispersion is $\sigma_R$ = 30 km/s and $\sigma_\theta$ = 20 km/s. This is the in-plane velocity dispersions for stars aged 4 Gyrs, as seen in \protect\cite{Nordstrom2004} or \protect\citep{Yu2018}. }
\label{fig:hits2}
\end{figure}

We have followed 400 000 stellar trajectories to see how often different orbits hit a GMC, and to remind the reader by this we mean when the star passes through the GMC, i.e when the distance between the star and the center of the GMC is smaller than the GMC radius. We find the number of hits depends strongly on the parameters $R$ and $V_z$ which can be seen in figure~\ref{fig:hits2}. Each line in figure~\ref{fig:hits2} represents trajectories with an average radius ($\left(R_{\rm max} + R_{\rm min}\right)/2$) binned to 0.5 kpc wide bins. $V_z$ is defined as the velocity away from the Galactic plane at $z = 0$. The decreasing number of hits with increasing $V_z$ is due to the fact that the GMCs are all located in a thin layer, as can be seen in table \ref{tab:naka}. With a higher $V_z$, a star spends less time close to the Galactic plane, as $V_z$ goes up and the vertical oscillation period of the star increases, resulting in less time spent in the GMC layer and fewer plane crossings per Gyr, lowering the probability of hitting a GMC. The $R$-dependence comes from two separate factors, first of all the surface density of GMCs goes down with increasing $R$ as can be seen in figure~\ref{fig:surf}. The second reason is that at larger $R$ a star gets to a larger $|z|$ for the same $V_z$ due to the shape of the potential.

We fit an exponential equation (equation \ref{eq:exp} below) to each line shown in figure~\ref{fig:hits2}. We find the lines to be well modelled by the equation and the resulting fits are shown in table \ref{tab:fits}.

\begin{equation}
N = A\exp\left(-\frac{V_z}{V_0}\right)+C
\label{eq:exp}
\end{equation}
where $N$ is the number of hits per Gyr and $C$ is a constant.

We then fit a single exponential to the number of hits as a function of radius for each $V_z$ and find that in this case the single exponential ($N=A\times\exp\left(-R/R_0\right)$) works well. We find the scale length to be 750 pc at low $V_z$, increasing to 850 pc as it reaches a peak at $V_z = 11$ km/s and then decreasing towards 800 pc.
\begin{table}
\centering
\caption{Fits using equation \protect\ref{eq:exp} which gives the mean number of GMC hits per Gyr at different radii as a function of $V_z$. The data which it is fitted to is shown in figure~\protect\ref{fig:hits2}. }
\begin{tabularx}{\columnwidth}{c c c c}
\hline
\small Radius (kpc) & A & $V_0$ (km/s) & C\\
\hline
\small 4.5  & 159.9$\pm$18 & 13.7$\pm$0.38 & 5.0$\pm$1.1\\
\small 5.0  & 127.3$\pm$12 & 12.3$\pm$0.29 & 4.7$\pm$0.7\\
\small 5.5  & 81.2$\pm$7.7 & 11.0$\pm$0.23 & 4.3$\pm$0.4\\
\small 6.0  & 55.0$\pm$3.4 & 10.9$\pm$0.14 & 2.7$\pm$0.16\\
\small 6.5  & 33.4$\pm$1.9 & 10.2$\pm$0.17 & 1.7$\pm$0.14\\
\small 7.0  & 19.1$\pm$1.1 & 9.3$\pm$0.21  & 1.2 $\pm$0.07\\
\small 7.5  & 12.8$\pm$0.8 & 9.1$\pm$0.19 & 0.7$\pm$0.06\\
\small 8.0  & 7.2$\pm$0.8 & 9.1$\pm$0.22 & 0.4$\pm$0.03\\
\small 8.5  & 4.5$\pm$0.6 & 9.1$\pm$0.25 & 0.3$\pm$0.02\\
\small 9.0  & 3.3$\pm$0.4 & 9.1$\pm$0.25 & 0.3$\pm$0.02\\
\small 9.5  & 2.6$\pm$0.4 & 9.4$\pm$0.31 & 0.3$\pm$0.02\\
\small 10.0  & 2.3$\pm$0.8 & 8.8$\pm$0.55 & 0.3$\pm$0.02\\
\hline
\end{tabularx}

\label{tab:fits}
\end{table}
\subsection{Distribution of the number of hits}
We look at how the number of hits are distributed at all radii within a single $V_z$ bin. In figure~\ref{fig:hithist} we see how the hits are distributed at all radii for $V_z=11$ km/s. 11 km/s is chosen because it is the velocity at which orbits start getting above the ${\rm H}_2$ layer (11 km/s gives a maximum z of $\sim50$ pc at 5 kpc). We see a fairly narrow distribution at small radii which widens significantly as the radius is increased. In figure~\ref{fig:hithist} we can see that 95\% of of the stars at 4-5 kpc have a number of GMC hits within a factor of two of the mean. However beyond 8 kpc we see larger deviations from the mean with a fraction having several times the mean in the number of hits and an even larger fraction ($\sim$20\%) with no hits at all during 1 Gyr. 

There might be very little difference between having 50 or 100 hits in a Gyr, it all depends on how affected the planet is by the perturbation and how fast it recovers. On the other hand, the difference between zero and three hits in a Gyr could be the difference that allows conditions for life be maintained sufficiently long for it to emerge.

\subsection{Kinematic heating, eccentricity and age}
We set the stars on non-circular orbits whilst we approximate GMCs to be on circular orbits (reasonable as they have a low velocity dispersion see e.g. \citealp{Gammie1991}). The stars being on non-circular orbits means that stellar orbits can cross GMC orbits as the stars move radially and in doing so they can collide. One might think that the more non-circular a stellar orbit is, the higher the probability is for collisions as it would cross more GMC orbits. To study the effects of non-circular orbits we define a Keplerian eccentricity for the stars as $e=(R_{\rm max}-R_{\rm min})/(R_{\rm max} + R_{\rm min})$ where $R_{\rm max}$ and $R_{\rm min}$ are the maximum and minimum radius reached during the simulation. We find that 95\% of the stars have  $e < 0.11$ and we find no correlation between $e$ and the number of GMC hits. In order to confirm the robustness of this result we do a set of runs in which we give the GMCs a small in-plane velocity dispersion (5 km/s in R and $\theta$) and we find that the results are the same as if we were to increase the dispersion of the stars. 

The result is intriguing because the eccentricity is a consequence of the kinematic heating a star experiences as it ages and along with increasing the eccentricity, the kinematic heating also increases the $V_z$. Since the eccentricity appears to have no effect on the number of GMC hits it means that we can keep the in-plane velocity dispersion constant, use a range of $V_z$ and in doing so test stars at different ages. \textit{Therefore, as a star ages its hit frequency with GMCs goes down}. Let us take a typical star in the Solar neighbourhood as an example. In the Solar neighbourhood the vertical age-velocity dispersion relation for stars 1-8 Gyrs old is $\sigma_{\rm z} \in [9.3, 23.5]~{\rm km/s}$ \citep{Nordstrom2004}. This $\sigma_z$ gives a mean $|V_z|\in [7.7, 19.4]~{\rm km/s}$. Using equation \ref{eq:exp} we see that a star with an average heating history in the Solar neighbourhood decreases its hit rate by a factor of 1.6 as it ages from 1 to 8 Gyrs.\\

\subsection{Radial migration}
The radial dependence on the number of hits per Gyr is strong. The strength of the gradient becomes particularly important when the radial migration of stars is considered, a process in which stars can change their guiding radius without becoming more eccentric due to interactions with spiral arms \citep{Sellwood2002}. The reason this is important is because more stars have been formed at 5 kpc compared to 8 kpc and perhaps more importantly, due to the Galaxy forming ``inside out" \cite{Martig2014} stars started forming earlier and with a higher metallicity at smaller radii. Some of these would have migrated outwards to a region in which the potentially habitable planets orbiting them are significantly less perturbed by GMCs, e.g. a migration from 5 to 8 kpc along with a heating in $V_z$ by 12 km/s implies the hit rate lowering by a factor of $\sim$100. The stars that formed early on with a high metallicty and migrated outwards when they were young are likely the first planets on which life would have developed in the Galaxy. 

\subsection{Trajectories with zero hits}
\begin{figure}
\centering
\includegraphics[width = 1\columnwidth]{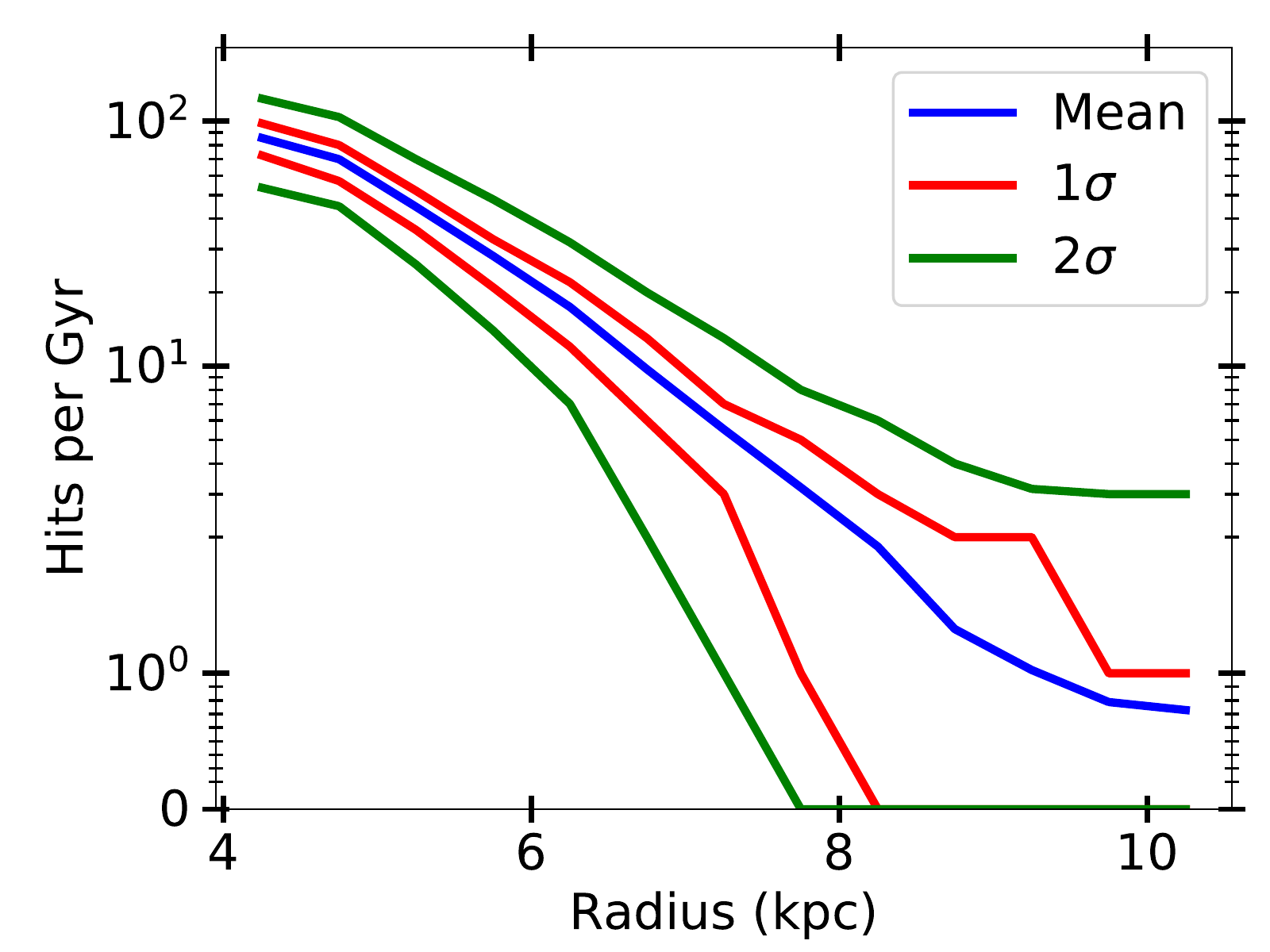}
\caption{The mean number of hits per Gyr in each mean radial bin with an initial $V_z=11$ km/s plotted together with the spread. The 1 and 2 $\sigma$ indicate the bounds where 68\% and 95\% of the data is contained, respectively.}
\label{fig:hithist}
\end{figure}

In figure~\ref{fig:hithist} we see that after a certain radius there is a fraction of trajectories that experience no GMC hits during 1 Gyr. A natural question to pose when seeing this is; are there trajectories that \textit{never} hit GMCs? Such trajectories would always go above/under the spiral arms and in doing so never hit the GMCs (perpetual missers). One can also make an intuitive argument against them, for perpetual missers to exist the azimuthal frequency must equal an integer times the vertical oscillation frequency because then if it misses all the arms during one revolution it will do so next time as well. As a star orbits it changes radius and therefore the azimuthal and vertical oscillation frequency changes. They do not change in the same manner and thus no perpetual missers can exist. To check this, we re-ran the missers for another Gyr and found that at a given $R$ and $V_z$ the same fraction of the missers remained missers as had been the original fraction of missers. Redoing this a few times we find that eventually all stars hit a GMC at inner radii ($<$ 8 kpc) where the GMC density is higher. At larger radii, there is a small subset of orbits ($\sim$1\%) that avoid hitting GMCs yet get scattered to such high $V_z$ that they can avoid hitting GMCs for 10 Gyrs, essentially making them perpetual missers. These kinds of orbits are discussed further in section~\ref{sec:sun}.

\subsection{Different spiral arm models}
As shown in table \ref{tab:models} we tested a few different models for the spiral arms to see if any difference in the number of hits was observed. For every model except E (in which the number of GMCs are doubled), there was no difference in the overall rates. For E it was as expected, doubling the number of GMCs doubled the number of GMC hits. However this isn't fully representative of what really happens since collisions between GMCs and their subsequent mergers were not considered which at this high a density becomes relevant. So to answer what the mass-scaling really looks like we need a more realistic treatment of the GMC dynamics. 

There was something interesting observed in model C (co-rotation at 6 kpc). Even though the mean number of hits for the trajectories were the same as in model A at all locations and for all $V_z$. When looking at the distribution in time for the hits at 6 kpc we found that nearly all hits happen at one or two intervals of time. This happens due to how our GMC trajectories are defined, near co-rotation the spiral arm will be much more narrow which makes it much more dense in GMCs. Once a stellar trajectory gets close to the arm it will stay there for a long time since the relative velocity is so low and get hit a lot of times. The same effect was not observed in Model B (co-rotation at 8 kpc), because there are so few hits for each trajectory at and around 8 kpc. The fact that nothing happens even when we artificially increase the density of the arms like this has implications for when discussing the periodicity of mass extinctions on Earth and ascribing it to spiral arm crossings. We find the same thing as \cite{Feng2013}; for most spiral arm crossings at 8 kpc nothing happens.

\section{Discussion}\label{sec:four}
\subsection{Putting supernovae in the GMCs}\label{sec:supernova}
An extension of our model is to also use it to calculate how often a star passes near a supernova as it explodes. The lifetime of stars that end their life as a core-collapse supernovae ($M_*\gtrsim 8{\rm M}_\odot$) is about the same as a GMC lifetime (see e.g.  \citealp{Zapartas2017}). This means that nearly all core core-collapse supernovae explode inside or near the GMC in which they are formed; this has been confirmed observationally by \cite{Gao2001}. For this reason we can calculate the probability of a star encountering a supernova when a GMC is hit, apply this to the simulated GMC hit history of our stars and find the probability of being within a certain distance of a supernova when it explodes.

We calculate the probability of a star being within 10 pc of a supernova when it explodes. The mechanisms in which a supernova can cause a mass extinction was discussed in the introduction, they are: direct irradiation \citep{Beech2011}, destruction of ozone layer and consequent death by Solar radiation \citep{Ruderman1974}, triggering an ice age by increasing albedo by forming NO${}_2$ in upper atmosphere \citep{Thomas2005} and triggering an ice age by having cosmic rays seed cloud formation \citep{Tanaka2006}. Which of these mechanisms will be dominant is unclear and it will likely depend on the supernova itself along with the environment in which it happens. The distances for which the supernova can cause mass extinctions is up to 15 pc, however these distances are derived assuming an isolated Sun. The supernova in our simulations go off inside GMCs and a star inside a GMC will be shielded by the gas. How much it is shielded will depend on what kind of GMC it occurs in and the location in the GMC. Due to this uncertainty we adopt a 10 pc kill radius in our experiment. From now on in this paper when we refer to being affected by a supernova we mean being within 10 pc of it when it explodes.

We do a Monte Carlo simulation in which we draw trajectories within a range of measured hit velocities randomly through GMCs over the range of masses and sizes included in the model. As the star passes through the GMCs supernovae are detonated and we count how many of them are within 10 pc and get the probability of a supernova affecting the star. A detailed description of the Monte Carlo simulation follows in the section below.

\subsubsection{Supernovae during GMC passages}

The number of supernovae per GMC is determined by the amount of gas converted to stars. We take all the GMCs to have a star forming efficiency of 10\%, consistent with recent observations such as  \cite{Ochsendorf2017}. Once we know how much gas is being made into stars we draw stars from a \cite{Kroupa2001} IMF shown below:
\begin{equation}
\frac{dN}{dM_*} = C_\alpha~M_*^\alpha
\end{equation}
\noindent where $C_\alpha$ is a constant dependent on $\alpha$ chosen such that the function is continuous and
\begin{equation}
\alpha =
\left\{\begin{array}{lr}
        -0.3, &   0.01<M_*/M_\odot<0.08 \\
        -1.3, &  0.08<M_*/M_\odot<0.5 \\
        -2.3, &   0.5<M_*/M_\odot<100
        \end{array}\right\}
\end{equation}
Once  stars have been drawn to reach 10\% of the total GMC mass, we stop drawing additional stars. Using the Kroupa IMF results in one in 300 stars being above 8 solar masses, or; for every 100 $M_\odot$ of stars formed we get one supernova. At this point we can do a reality check of our model by calculating what star formation rate we end up with under our assumptions and comparing it to observations. Taking a total ${\rm H}_2$ mass of $6.5\times10^8{\rm M}_\odot$ with 40 Myr lifetime of GMCs during which 10\% of their mass is converted to stars we end up with a SFR of 1.64 $M_\odot~{\rm yr}^{-1}$. This agrees well with observations; \cite{Licquia2015} find a SFR of $1.65\pm 0.19 {\rm M}_\odot {\rm yr}^{-1}$.

The supernovae are spatially distributed within the GMC in a fractal manner following the prescription by \cite{Goodwin2004}. To each supernova we then assign a detonation time, determined by their mass with the delay time distribution given in \cite{Zapartas2017}. We treat all the star formation in the GMC as occurring instantly when the GMC is 10 Myrs old.

The fractal distribution is generated by taking a cube and splitting it evenly into 8 sub-cubes. A sub-cube is accepted with probability $P = 2^{{\rm D}-3}$. If accepted, the cube is randomly nudged in a direction to avoid having them all on a lattice and then split further into another 8 sub-cubes who are also accepted with probability $P = 2^{{\rm D}-3}$. Where D is the fractal dimension and D = 3 gives a homogeneous cube, we use D = 2 which is a reasonable dimensionality for open clusters \citep{Parker2014}. Once a generation of cubes whose number exceeds the number of supernovae required is generated, the cube is pruned into a sphere and a number of sub-cubes are removed to get the specified number of them. If the pruning removes too many sub-cubes the process is repeated. Once this is done, a supernova is put into the center of each remaining sub-cube.

We test a large range of relative velocities, we generate trajectories with velocities of 1-200 km/s isotropically crossing through the GMC, averaging over the impact parameters observed in the simulation. We then draw a time from a randomly uniform distribution between 0 and 40 Myrs to also average over potential ages. Then we check how often the trajectory is within 10 pc of a supernova as it explodes. We treat the expansion of the ejecta and the speed of light as infinite which is a reasonable approximation given typical relative speeds. This generates a function of GMC mass and crossing velocity giving an average number of supernovae per crossing. The average number of hits go up with GMC mass and go down crossing velocity. The decrease with velocity is non-linear as can be seen in figure~\ref{fig:snevel}. The average number of supernovae a star is affected by does not go up sharply with GMC mass as one might have expected. This because the GMC density goes down with increasing mass as one can see in equations \ref{eq:mr}-\ref{eq:mr3}, resulting in the supernova distribution getting sparser. 
\subsubsection{Supernova results}
\begin{figure}
\centering
\includegraphics[width=1\linewidth]{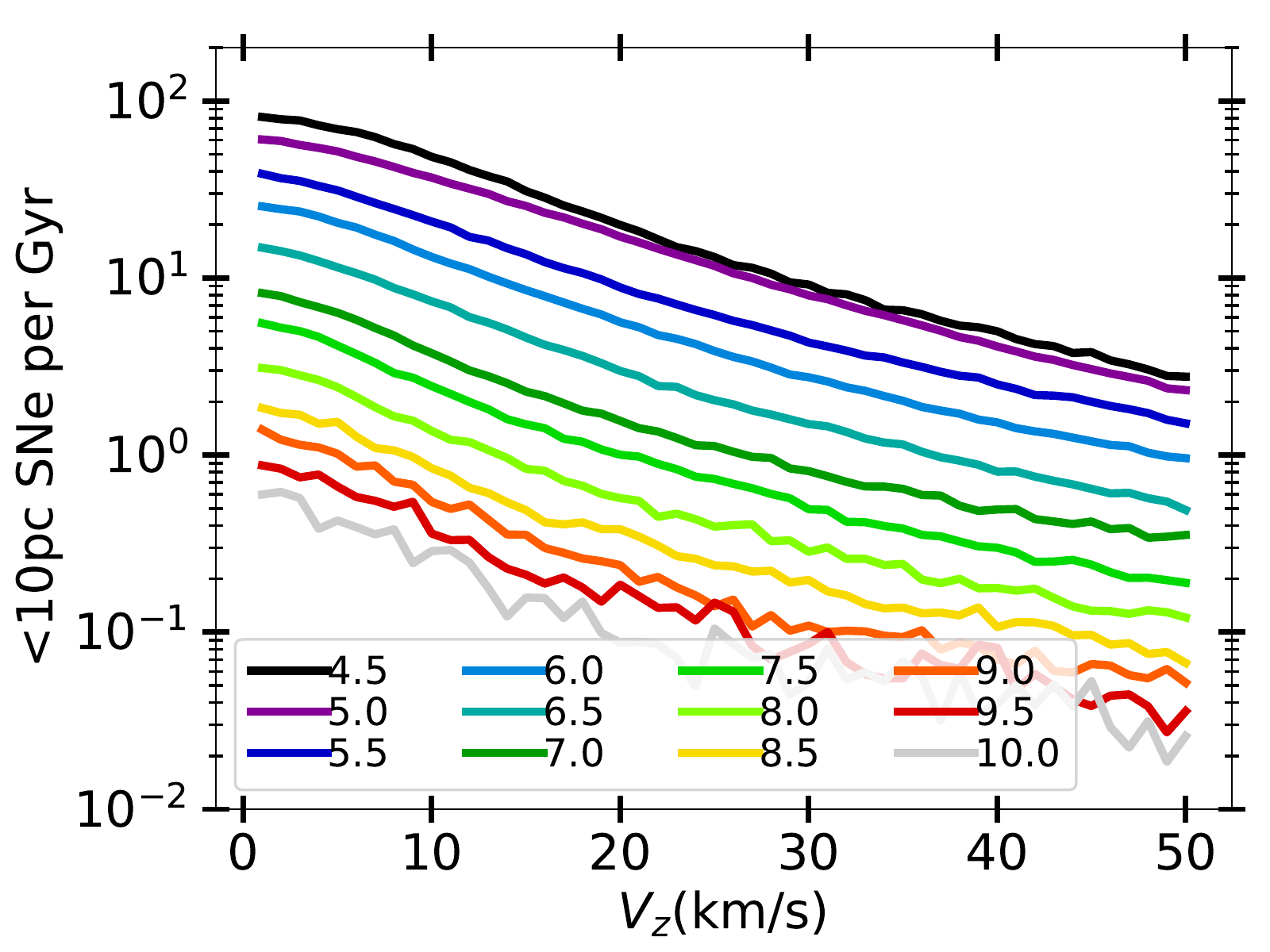}
\caption{The average number of $<10$ pc supernovae affecting a star during 1 Gyr as a function of initial $V_z$. The figure was produced by taking the results of the procedure described in section \ref{sec:supernova} and applying it to the data plotted in figure~\ref{fig:hits2}. }
\label{fig:SNresult}
\end{figure}

\begin{figure}
\centering
\includegraphics[width=1\linewidth]{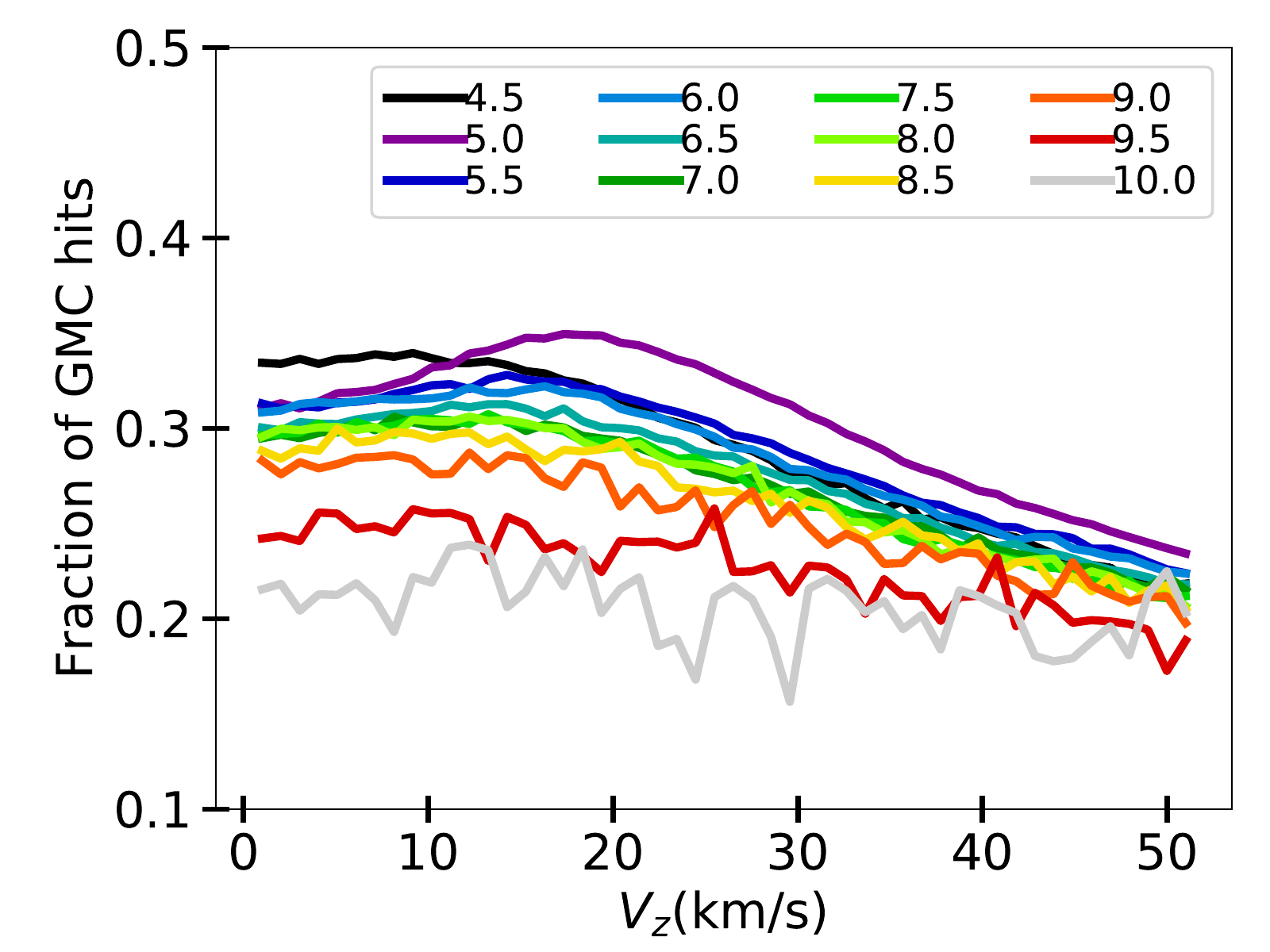}
\caption{The fraction of GMC hits at different mean radii and initial $V_z$ that result in at least one supernova affecting a star within 1 Gyr.}
\label{fig:SNfrac}
\end{figure}

Having determined the average number of supernovae a star is affected by when a GMC is hit we can take the masses and impact velocities from the trajectories described in the previous section and calculate how many supernovae they would have been affected by during 1 Gyr. Figure~\ref{fig:SNresult} shows how many supernovae a star is affected by  during 1 Gyr for different trajectories. The results look similar to what we see in figure~\ref{fig:hithist}, however there is one key difference. The fall-off with $V_z$ is larger because not only does it reduce the probability of hitting a GMC, but it also increases the velocity relative to the GMCs and in doing so it reduces the probability of being affected by a supernova because the star spends less time within 10 pc of the potential supernova. It is important to note that quite a bit of information is lost when using the mean because the distribution is quite lopsided. In the majority of GMC crossings the star is not affected by any supernovae. Due to the fractal and lumpy distribution of the supernovae not only does the trajectory have to cross through the GMC at the right time, but also in the right place. When this happens the star is affected by several supernovae per crossing. So we can ask a different question; when a GMC is hit, what is the risk of being affected by at least one supernova? We take the distributions calculated in the previous subsection but now when determining the function we don't allow more than 1 supernova per crossing. The resulting risk of being affected by a supernova is shown in figure~\ref{fig:SNfrac}. In it we see what we expect in that the risk of being affected by a supernova goes down with $V_z$ since the relative velocity increases and we also see that the risk ranging between 0.2-0.3. If we just take the number of supernova affecting the star per GMC hits we find it to be 0.4-0.5. 

\begin{figure}
\centering
\includegraphics[width = 1\columnwidth]{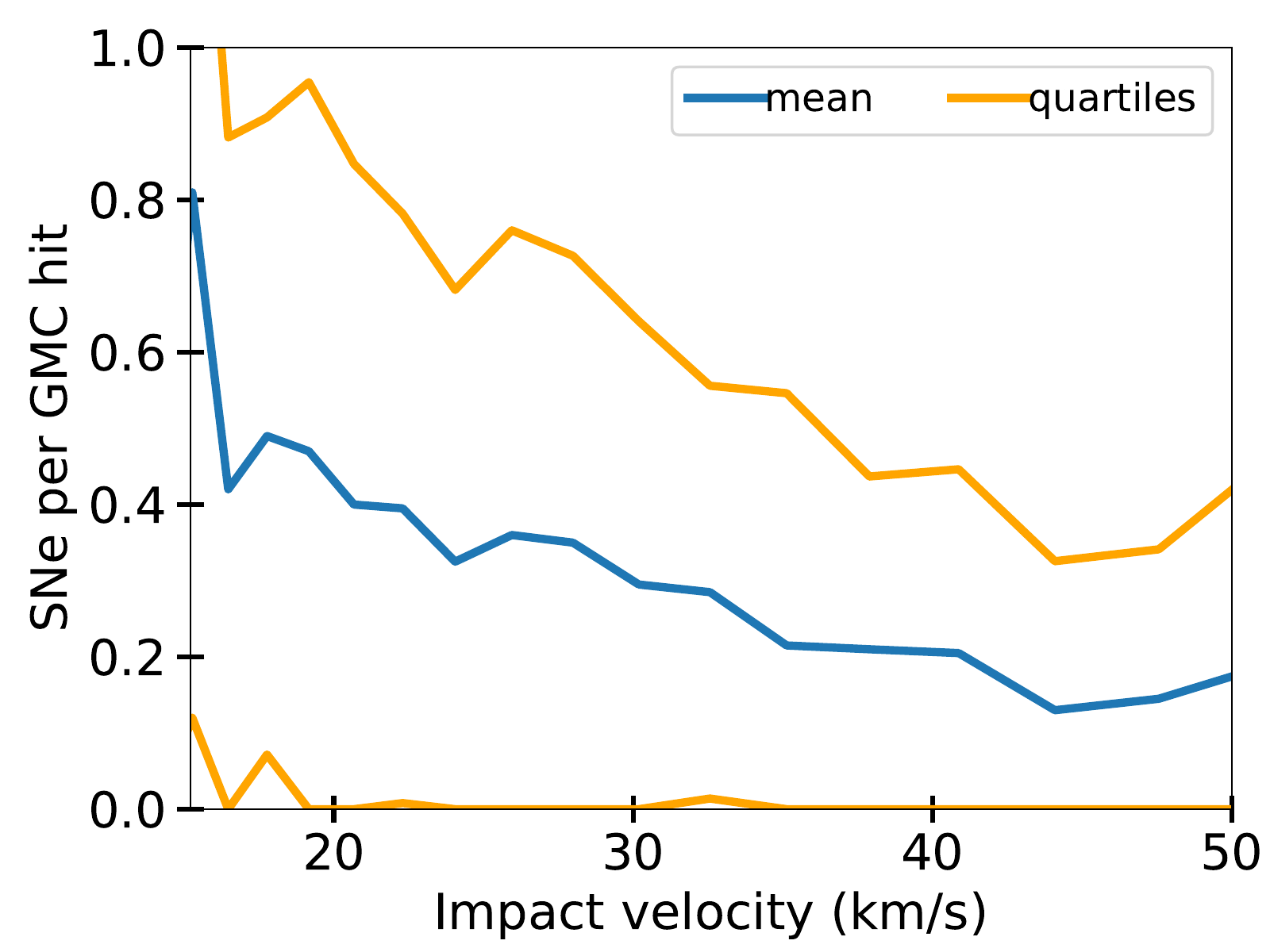}
\caption{The average number of supernovae a trajectory is affected by when hitting the most common type of GMC, i.e. a $8\times10^5 {\rm M}_\odot$ dense GMC. The x-axis starts from the escape velocity of the cloud as slower hits than that are not possible and the noise in the function is due to the inherently stochastic process in which it is generated.}
\label{fig:snevel}
\end{figure}

The median GMC hit mass is large at $8\times10^5{\rm M}_\odot$, this has a consequence for the probability of being affected by a supernova. The large mass means that the star accelerated from a few up to almost 10-15 km/s depending on the geometry and initial relative velocity of the GMC hit. The resulting effect is a quicker passage through the GMC. How many supernovae the star is affected by during a passage is a dependent on velocity as can be seen in figure~\ref{fig:snevel}. So an increase in velocity by $\sim$10 km/s can significantly lower the probability of being affected by a supernova. This means the probability of being affected is enhanced in lower mass GMCs, compensating for the fact that they produce fewer supernovae.

\subsection{Testing with Solar parameters}\label{sec:sun}
To find the rates for the Sun, we use the Solar orbital parameters as determined by \cite{Dehnen1998} to initialize a set of runs. The orbital parameters given as U, V, W are the peculiar velocities of the Sun defined as; the deviation from the local standard of rest along the Solar orbit, radially inwards to the Galactic centre and away from the Galactic plane towards the position of Sun respectively. They are  [$(U,~V,~W)$ = (10.0$\pm$0.35, 5.23$\pm$0.65, 7.17$\pm$0.38) km/s], they are lower than an average 4.5 Gyr old star for which the average peculiar velocities are (23, 16, 11) km/s.  \citep{Nordstrom2004}. Whilst the non-circular component of the velocity has little to no effect on the hit rate of GMCs, it will affect the supernova probability as it increases the relative velocity of the hit. 

\begin{figure}
\centering
\includegraphics[width = 1\columnwidth]{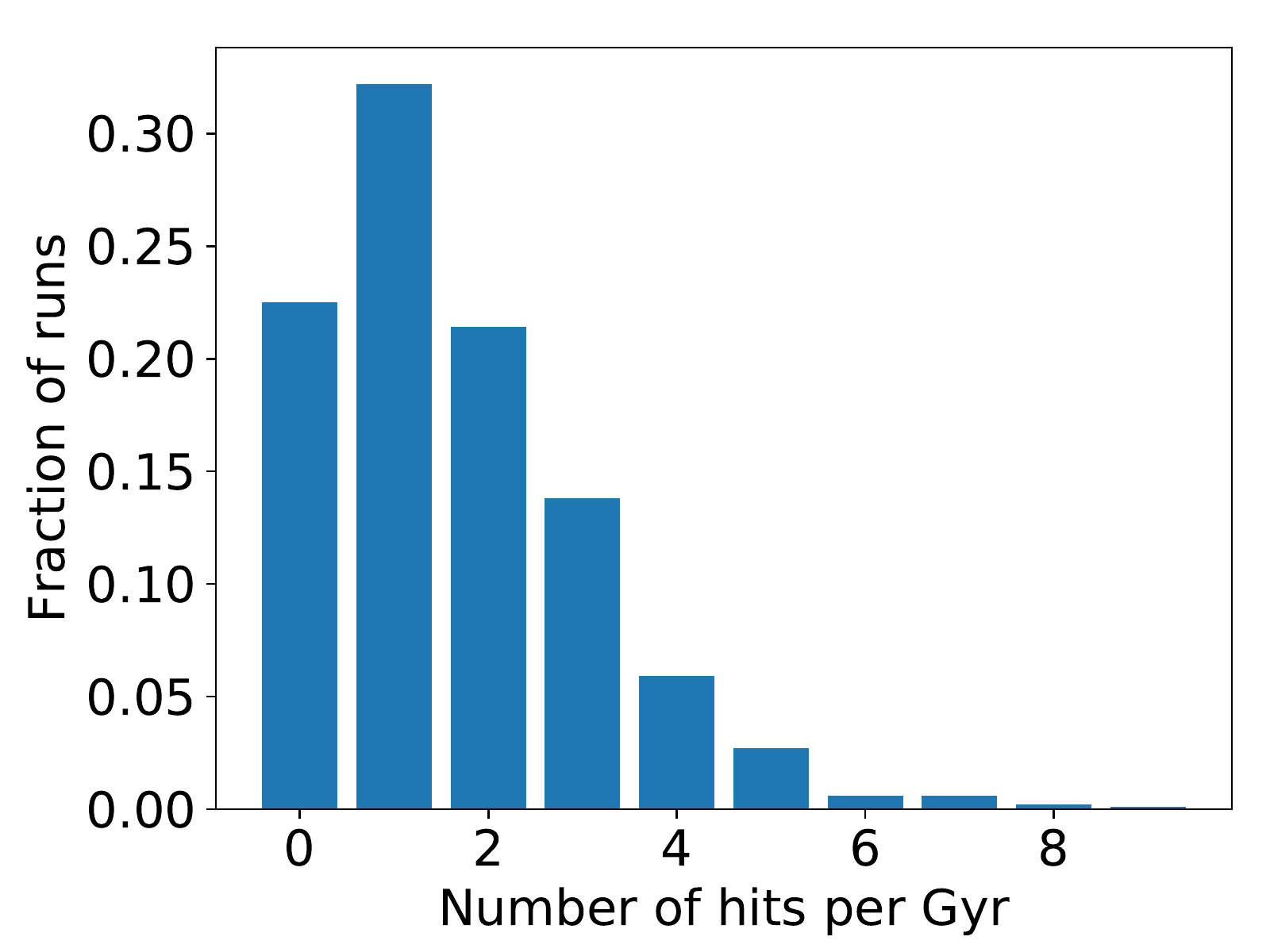}
\caption{The number of hits per Gyr for a set of ``Sun-like" runs. I.e. starting the trajectory at 8 kpc and giving it the non-circular velocity components from \protect\cite{Dehnen1998}. The difference between each run being the angle at which the trajectory starts}
\label{fig:sunhits}
\end{figure}

Using the Solar values of peculiar velocities we initialize 1000 trajectories at randomly selected $\theta$ at 8 kpc, and the velocities are given small differences that are drawn from the Gaussian errors shown above. Then we see how often a Sun-like trajectory will hit a GMC and be affected by a supernova. Figure~\ref{fig:sunhits} shows the results from using a Sun-like orbit. In it we see that the Sun will hit an average of $1.6\pm1.3$ GMCs per Gyr, or, one GMC every $625^{+2700}_{-280}$ Myrs. The deviation in $1.6\pm1.3$ comes from fitting a Poisson distribution to the number of hits per Gyr, seeing it matches and then taking the Possion standard deviation; $\sqrt{\lambda}$ where $\lambda$ is the mean. Then $625^{+2700}_{-280}$ Myrs is simply its inverse. We could apply the supernova function from section \ref{sec:supernova} directly, however due to the Sun's low peculiar velocity it passes through GMCs much more slowly. Therefore it is affected by at least one supernova during nearly half (47\%) of the hits. It should be noted that due to the slow hits the Sun is even more prone to multiple supernova per GMC hit, if we were to look at the average rather than the risk we find 0.9 supernova affecting the Sun on per GMC hit. This means that when the Sun experiences a nearby supernova it more often than not experiences multiple supernovae in a short time-period. So if we just look at the average we find that the supernova occur once every $667^{+2500}_{-240}$ Myrs, however if we correct for the spatial and temporal clustering by assuming that two supernovae within a few Myrs is as bad for habitability as one we instead find supernova occurring once every $1.25^{+5.25}_{-0.55}$ Gyrs. 

\begin{figure}
\centering
\includegraphics[width = 1\columnwidth]{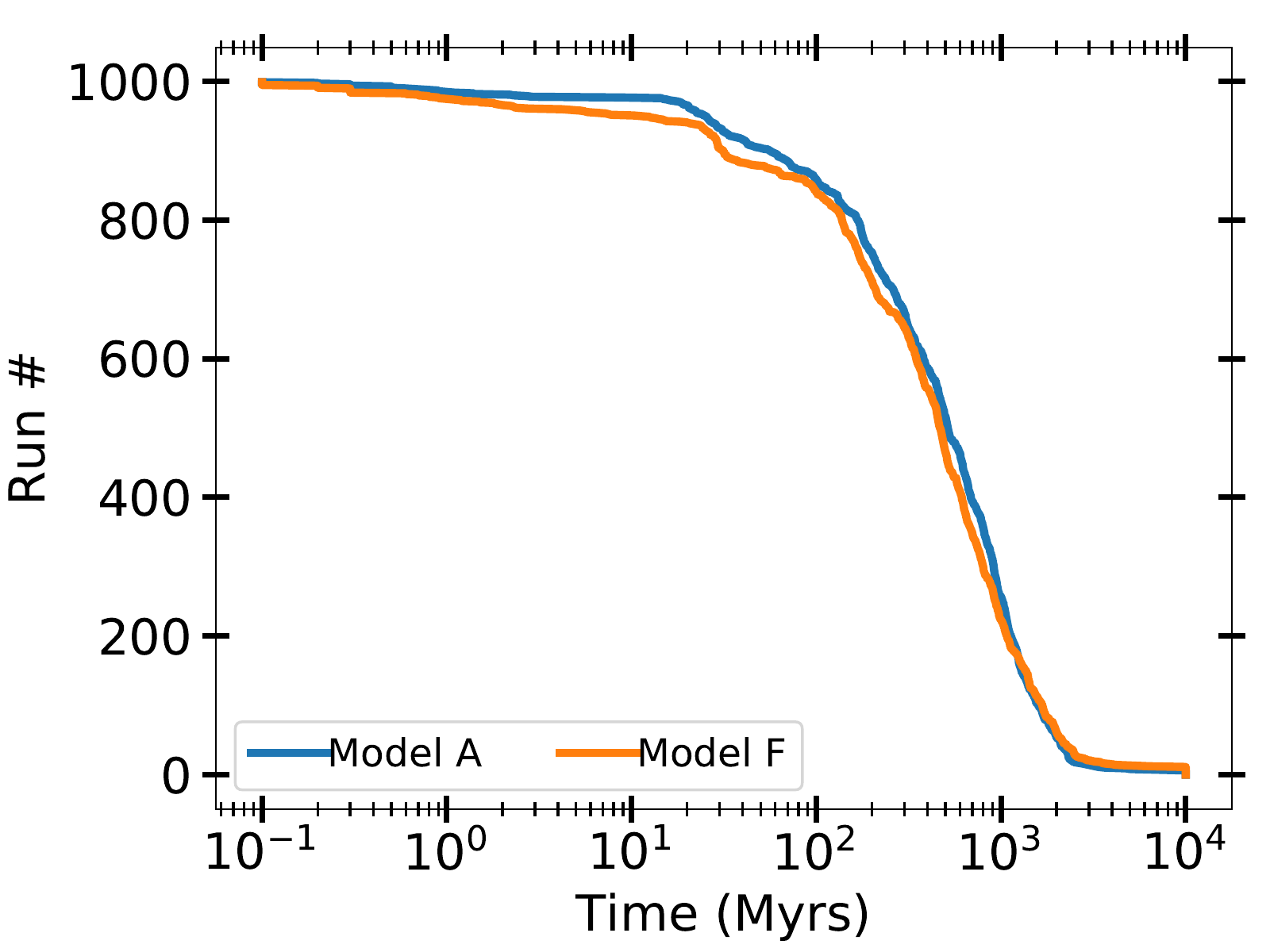}
\caption{The time until the first GMC hit for each Sun-like trajectory in an extended, 10 Gyr, simulation. In the bottom there is a subset of runs that never hit a GMC, this is because they get scattered to such a high $V_z$ that the hit probability gets sufficiently low for them to avoid hits for the full 10 Gyrs. }
\label{fig:timehit}
\end{figure}

However, there is a tail to the distribution that doesn't arise from just the stochasticity of the hit probabilities. We looked closer at this by doing a 10 Gyr integration for the Sun-like runs. Figure~\ref{fig:timehit} shows the time until the first GMC hit for all the Sun-like runs, in it we see the tail ($\sim$1\% of the runs), these are runs that get scattered to high $V_z$ without actually hitting a GMC and in doing so it makes it possible for them to avoid hits for $>$5 Gyrs. In the figure we see two lines, one for model A and one for model E. This further exemplifies that the spiral arm model does not matter for the results. 

Our average time between nearby supernovae affecting the Sun differs from some of the reported values (such as  \cite{Ellis1995} - 243 Myrs, \cite{Tanaka2006} - 100 Myrs or \cite{Sørensen2017} - 183 Myrs ). The primary reasons for the difference is that they use the radial distribution of stars or the local stellar density in their calculation which we argue below is not an accurate way of calculating the probability of supernova occurring nearby. Supernovae in our Galaxy model follow the distribution of the GMCs, because that is where star formation and thus SNII occur. The radial surface density distribution of GMCs and stars are different. The surface density of stars follow an exponential with a $2.6\pm0.6$ kpc scale length \citep{Bland-hawthorn2016}. The molecular gas density decreases rapidly at first, then it plateaus between 3-5 kpc and then continues rapidly to fall off (as seen in figure~\ref{fig:surf}). The fall-off after the plateau we find to follow an exponential with a scale length of 1.73 kpc in our simulation with data from \cite{Nakanishi2006}. One can fit an exponential to the ${\rm H}_2$ distribution as a whole and then one ends up with a scale length of $\sim2$ kpc \citep{Miville-Deschenes2017}. Naively one might assume that the two distributions should follow each other as stars form from the gas, however stars are on average moved outwards to larger radii due to radial migration whereas the tendency for the gas is the opposite \citep{Sellwood2002}.

\begin{figure}
\centering
\includegraphics[width = 1\columnwidth]{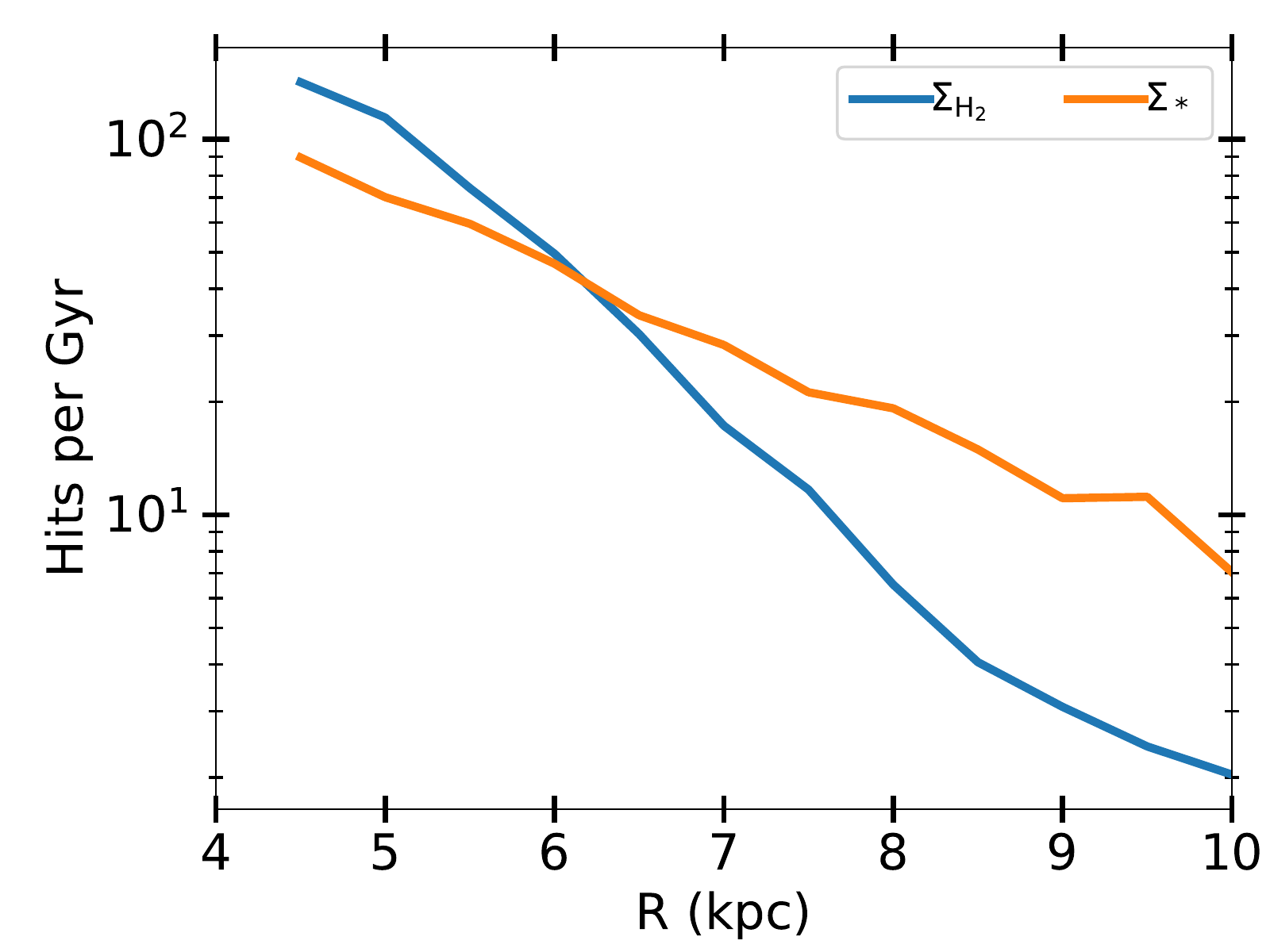}
\caption{The number of hits as a function of radius for stars with initial $V_z$ = 1 km/s. The blue line shows the result when using the radial distribution of the molecular gas whereas the orange line shows the result for when using a stellar exponential radial distribution. We see here that the gas and therefore the SNII are more centrally concentrated than the stars and therefore the SNIa. We chose to do the comparison at the lowest $V_z$ we test for since when local supernovae rates are calculated, motion in the z-direction often is ignored.}
\label{fig:expR}
\end{figure}

We investigate this difference by distributing the GMCs following the stellar surface density and then redo the simulation with $\sigma_R$ = 25 km/s and $\sigma_\theta$ = 15 km/s. Figure~\ref{fig:expR} shows the resulting number hits for a nearly flat orbit ($V_z$ = 1 km/s) using the two different radial distributions of GMCs. The figure shows a higher hit rate for the gas distribution further in which then crosses over at 6 kpc and has a lower hit rate further out, resulting in a factor of $\sim4$ difference at the Solar circle. The supernovae being more centrally concentrated is also consistent with recent observational findings by \cite{Green2015}. 

It should also be noted that our Galactic supernova rate is lower than the previously mentioned work for two reasons. First of all our model is limited to looking at core-collapse supernova. Second, is that our supernova rate is more in line with up to date estimates of the Galactic supernova rate \citep{Li2011}. 

Nearby SNIa could be more frequent than core-collapse supernova as they follow the stellar distribution and make up 25\% \citep{Li2011} of the Galactic supernovae. We can estimate the time between nearby SNIa by taking the numbers derived in other work in which nearby supernovae are determined by looking at stars and multiplying them by four. This gives 732, 400 and 972 Myrs. Those numbers are very similar to and well within the uncertainty range of our first result (664 Myrs) and all of them are smaller than our second one (1.28 Gyrs). If we consider the assertion of \cite{Gowanlock2011} that SNIa are disruptive for planets they would certainly be the bigger threat regardless of model. That assertion was based on the absolute magnitude of the different class of supernovae and as we have discussed previously it is the FUV/X-ray luminosity that matters. Observations by \textit{Swift} \citep{Immler2006, Modjaz2009} indicate that even though SNIa are visually brighter than core-collapse supernovae, their luminosity in the shorter wavelengths are comparable and even pointing to the core-collapse being slightly brighter. For that reason we believe the risk posed by the different kinds of supernovae to be equal. 

\subsubsection{Matching the Geological record}
\label{sec:gr}
There is geological evidence of two supernovae going off within 30-100 pc of the Sun 1.5-3.2 and 6.5-8.7 million years ago \citep{Knie2004, Wallner2016, Statement2017, Hyde2017}. This is determined by looking at deep sea cores, where one finds deposits of ${}^{60}$Fe which can only exist on Earth if newly formed as it has a 2.6 Myr half-life. The conclusion is that it comes from supernovae and by looking at the amount of ${}^{60}$Fe one can estimate the distance to the supernovae.

Our model should be able to reproduce the geological record in one of two ways, either by encountering two separate GMCs between the arms and being hit by a supernova from each of them. Or, by being hit by two separate supernovae from a single GMC. From the GMC maps of the spiral arms in \cite{Hou2014} we know that the Sun is currently between spiral arms, and in our model we don't have any GMCs there. To be able to draw conclusions about the Sun's current location in the Galaxy we use both model A and F (table~\ref{tab:models}) with a key difference. The difference is that we do two runs, one in which we check when the trajectory gets within 100 pc (or is inside) of any GMC and one in which we check when the trajectory gets within 30 pc (or is inside) any GMC. Then given the encounter history, we calculate the probability of encountering two supernovae within 5 Myr of each other. With a caveat; the dating in the geological data has a $\sim1$ Myr uncertainty, therefore we stipulate that the supernovae hits on passage have to take place with at least 1 Myr separation in time. Neither of the GMC distributions represent the current GMC distribution around the Sun, however by interpolating between what we see in the different models we can draw conclusions about the Sun. To find the supernova history we redo the procedure described in the previous section but rather than using a range of masses and velocities, we use the ones we find in the simulation. Also, instead drawing trajectories through the GMC we draw them through 30/100 pc spheres centred on the GMC if the GMC is smaller.

\begin{figure}
\centering
\includegraphics[width = 1\columnwidth]{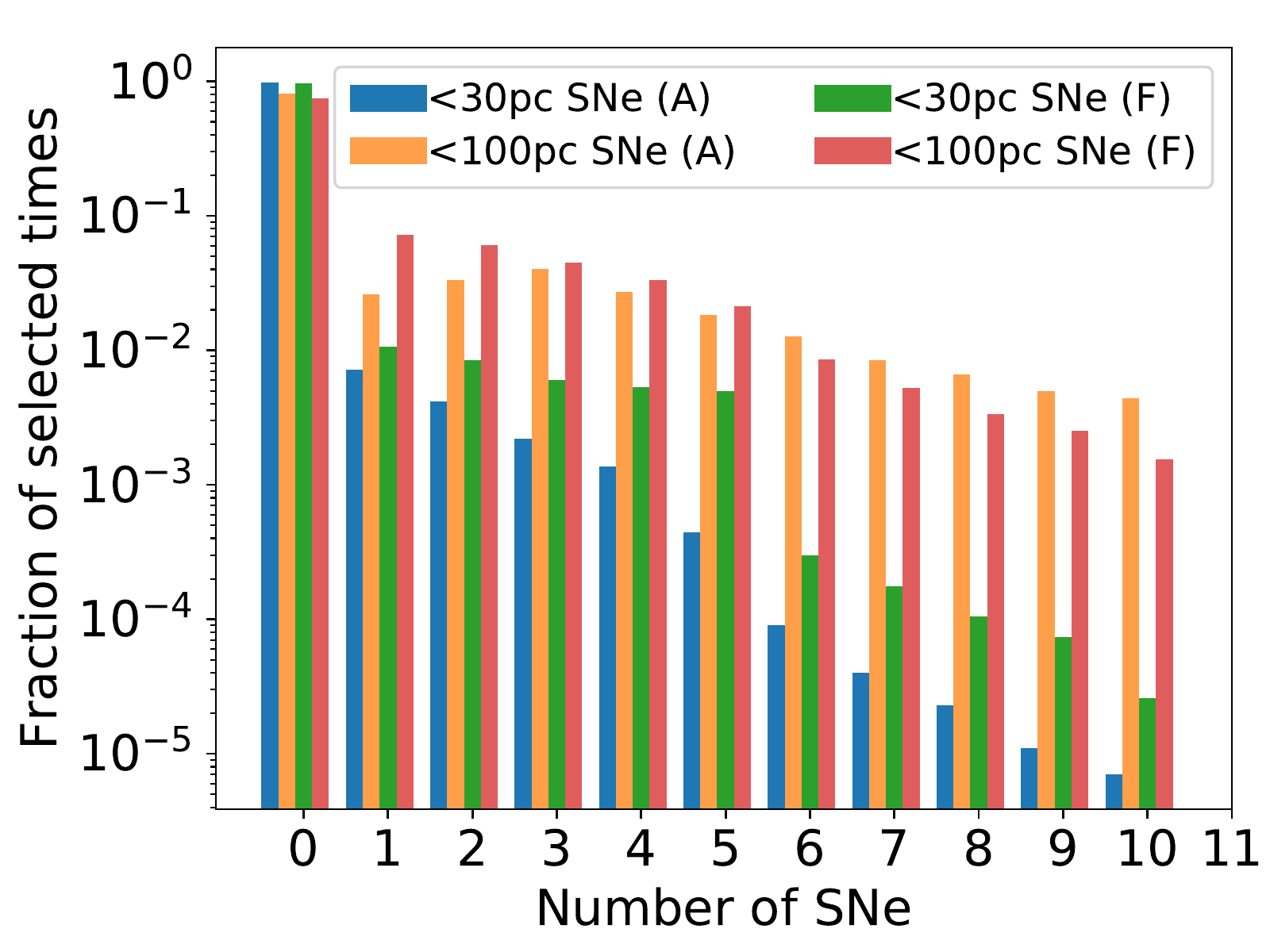}
\caption{The distribution of the number of supernovae found when randomly picking a 5 Myr interval. In it the result using models A and F (table \ref{tab:models}) are shown. }
\label{fig:distsupernova}
\end{figure}

In figure~\ref{fig:distsupernova} the number of supernovae one finds if one looks at a randomly chosen 5 Myr timespan for the 30 and 100 pc runs in the different models is shown. We see that it is $8\pm1.3$ and $6.9\pm0.8$ times more likely to have come from the 100 pc case in models A and F respectively. Which is slightly less than what one would expect from simple geometry. What we also see in Model A is an over-production of supernova, it becomes 32\% more likely to have three supernovae rather than two in the 5 Myr timespan. This is likely due to having 37\% of cases have supernovae from more than one GMC with model A compared to 24\% with model F. This means that our model reproduces the geological record if the Sun is currently in an environment with a higher GMC density than randomly spread in a disk, but lower than that of the spiral arm. That is consistent with being a spur and therefore the observations, and as a consequence of this we can say that our model is consistent with the the geological record and the observed spiral structure around the Sun. As for the distance to the supernovae it is clear that we favour more distant supernovae, however it must be noted that if we relax our one supernova per Myr criterion we see a higher multiplicity in supernova encounters per passage. This means that two distant supernova can easily go off within such a time-span as to be indistinguishable from a single, nearer supernova. For this reason, along with the low percentages of supernovae we see from different GMCs we favour the two supernovae having come from the same star forming region. The fact that it is not likely overall to have any supernovae during the 5 Myr interval is not a problem as we randomly draw from all times in figure~\ref{fig:distsupernova}, this includes times when not located near spiral arms. Since the Sun is currently located in the local spur we can disregard the high probability of no supernovae. 

Additionally, we also tested model B (corotation at 8 kpc). When at corotation and whilst being located in the local arm/spur we see much higher multiplicity of supernovae. If one supernova is observed we expect to see an order of magnitude more supernovae in the following 5 Myr compared to the homogeneous case. Therefore, we claim that it is unlikely we are at corotation as suggested by \cite{Junqueira2017}.

\noindent However, it should be stressed that to fully understand the history of the Sun it is necessary to investigate all aspects and in particular to utilize the geological record when possible such as the record presented in e.g. \cite{Shaviv2003, Shaviv2014} where they look at cosmic ray tracers in iron meteorites and fossils, respectively. This is however outside of the scope of this paper, but can be explored in future work as an extension of our model.

\subsubsection{Stripping the Oort Cloud}
As the Sun passes through a GMC it will feel the tidal force being exerted upon it. The Oort cloud \citep{Oort1950}, which is a cloud of comets surrounding the Sun extending out to $\sim$1 pc, is due to its size very susceptible to tidal disruption. Previous work on Oort cloud depletion due to tidal interactions have been done, e.g. \cite{Napier1982, Hut1985}. These two works use different methods and arrive at very different survival fractions of the Oort cloud, for this reason we will look closer at them and see how their results would have changed given the results from our model of the GMC hit history. 

\cite{Napier1982} calculate their depletion by integrating a set of Solar trajectories as they pass through a GMC. Where they use $5\times10^5M_\odot$ as a typical GMC mass where the GMC consists of 25 sub-particles of $2\times10^4M_\odot$. They surround their Sun with 33 000 particles representing the Oort cloud and count what fraction of them are stripped away at each passage. The fraction removed is calculated for a range of velocities, which means that we can take the fractions and interpolate between them to calculate the depletion given our hit history. We have different encounter histories for the lifetime of the Sun (4.5 Gyrs). They assume 14 GMC hits, and we find the average number of hits during 4.5 Gyrs to be 7. They assume a typical encounter velocity of 20 km/s whilst in our runs we determine the typical velocity to be 15 km/s. As an approximation when using their results to calculate the survival fraction we will use their cloud mass ($5\times10^5 M_\odot$) which is slightly lower than the typical cloud mass encountered in our simulation ($8\times10^5M_\odot$) since it will not affect the outcome significantly. Since slower encounters remove a larger fraction of comets we end up with a similar survival fraction to theirs despite having only half as many encounters; we find a survival fraction of $2\times10^{-3}$ compared to theirs of $1.5\times10^{-3}$. The depletion rate is proportional to the number of comets at a given time so to calculate the survival fraction the depletion rate is integrated over time. For this reason we can, much like with radioactive decay, calculate a half-life for the Oort cloud. Doing that we find that a survival fraction of $1.5\times10^{-3}$ corresponds to a half-life of 480 Myrs and $2\times10^{-3}$ corresponds to a half-life of 510 Myrs.

\cite{Hut1985} do an analytical calculation of the Oort cloud depletion by GMCs. The work is based on their previous work on disruption of wide binaries by GMCs in the Galactic field \citep{Bahcall1985} in which they derive an equation for the half-life of binaries due to GMC interaction. The same equation is used (\citealp{Hut1985}, eq. 7) to estimate the half-life of the Oort cloud. They estimate the half-life to be 2.7 Gyrs. We do the same thing with \cite{Hut1985} and use their method with numbers from our model and from newer observations. The estimation is dependent on seven factors, four of which we can extract from observations and our simulation ($\rho_0,~f_z,~f_e,~f_p$ in eq. 7) and the remaining three are all dependent on powers of the velocity dispersion of the encounters. The velocity dispersion we find is 15 km/s, using that instead of their value of 22 km/s results in a half-life of 680 Myrs (survival fraction of $10^{-2}$).

\begin{figure}
\centering
\includegraphics[width = 1\columnwidth]{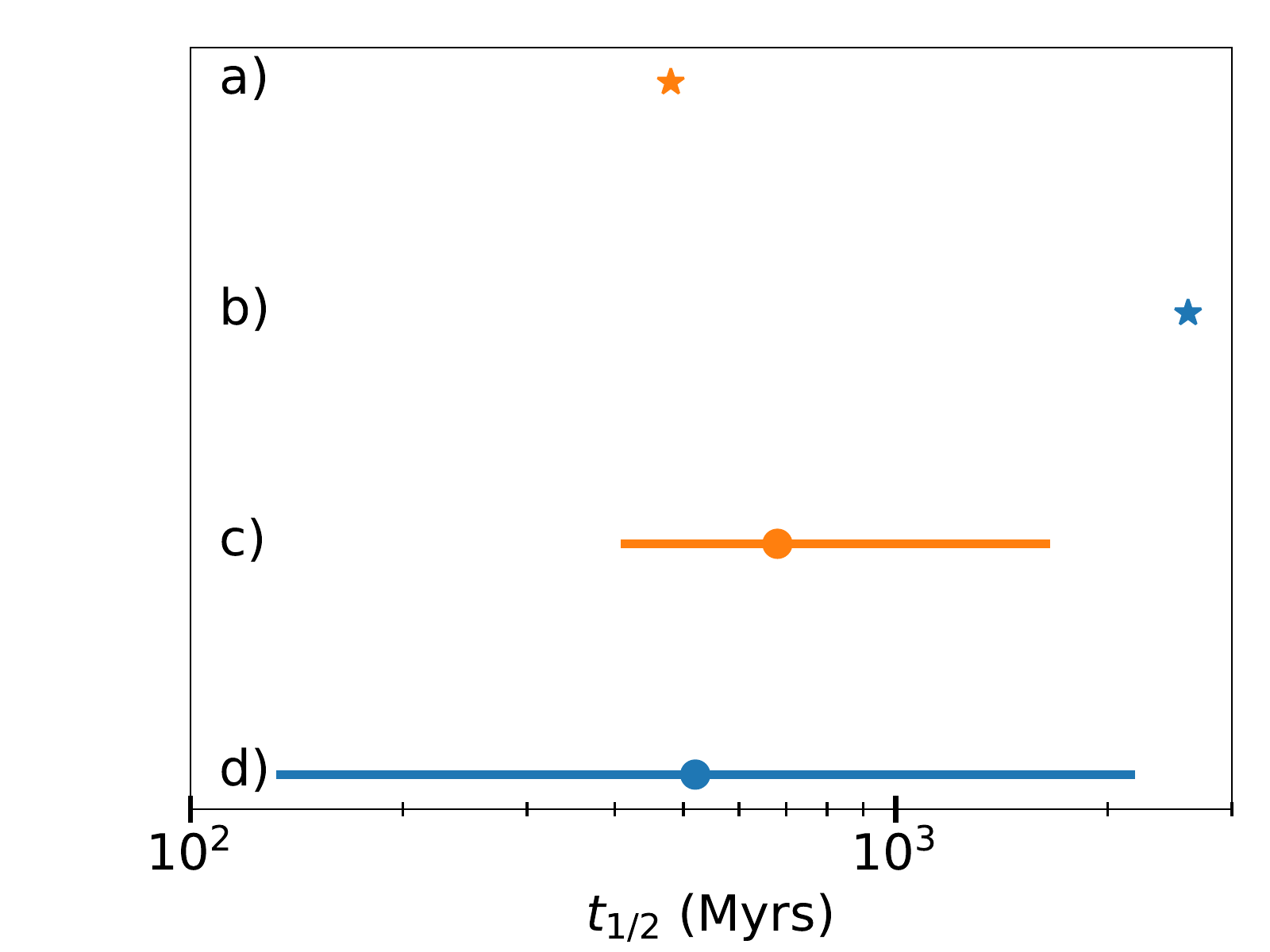}
\caption{The half-life of the Oort cloud computed in four different ways: \textit{a)} The half-life of the Oort cloud from \protect\cite{Napier1982}\textit{ b)} The half-life of the Oort cloud from \protect\cite{Hut1985} \textit{c)} The mean half-life and the 80\% interval calculated by taking the distribution of the number of GMC hits and hit velocities in our model and then interpolating in the results of \protect\cite{Napier1982} \textit{d)} The mean half-life and the 80\% interval calculated by taking the the distribution of GMC hit velocities in our model and applying those together with new observations to equation 7 in \protect\cite{Hut1985}.  } 
\label{fig:oclife}
\end{figure}

The half-lives calculated, using our numbers and applying it to their methods, are similar. This does not however mean that it is the definitive answer of how depleted the Oort cloud is; what this simple exercise has shown is how sensitive the depletion fraction is to the relative velocity of the encounters. The velocities we find have a wide distribution and we have also shown that the number of encounters vary enormously; taking these two facts into account we can derive the ranges of halkf-lives shown in figure \ref{fig:oclife}. Figure \ref{fig:oclife} implies that some Sun-like stars will have nearly pristine Oort clouds (few and fast encounters giving a long half-life) and some have very depleted Oort clouds (many slow encounters giving a short half-life). This has implications for both the past and future of these systems, the ones with pristine Oort clouds will have had few injections of comets in the past with a large risk of comet injection in the future whereas the opposite is true for the ones with a depleted Oort cloud. 

Additional uncertainties arise when considering that stars will also disrupt the Oort cloud. Recently, \cite{Hanse2018} looked at how stellar encounters strip the Oort cloud throughout the lifetime of the Sun. They simulate different orbital histories for the Sun and extract the resulting close encounter histories. With the encounter history they then simulate the single encounters and look at the exchange and loss of comets. They find that 25\%-65\% of the initial mass has been lost from just stellar encounters. 


\subsubsection{Accretion}
\begin{figure}
\centering
\includegraphics[width = 1\columnwidth]{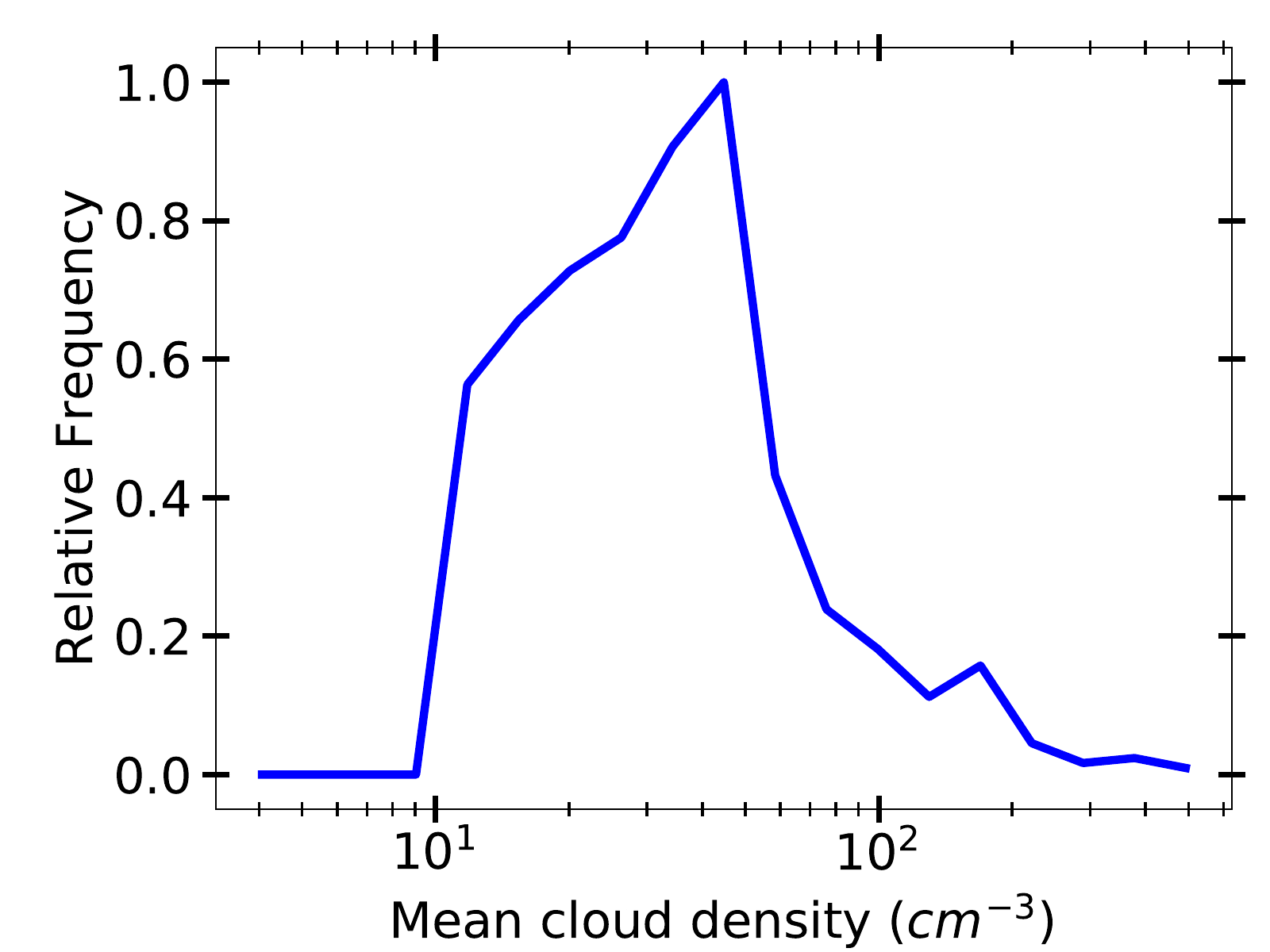}
\caption{The mean densities of clouds hit by a star on a Sun-like orbit. These are the densities used to calculate the accretion rates in figure~\ref{fig:accs}. The jagged corners in the distribution arise due to the mass-radius relationships defined in equations \ref{eq:mr}-\ref{eq:mr3}.}
\label{fig:rhos}
\end{figure}

\begin{figure}
\centering
\includegraphics[width = 1\columnwidth]{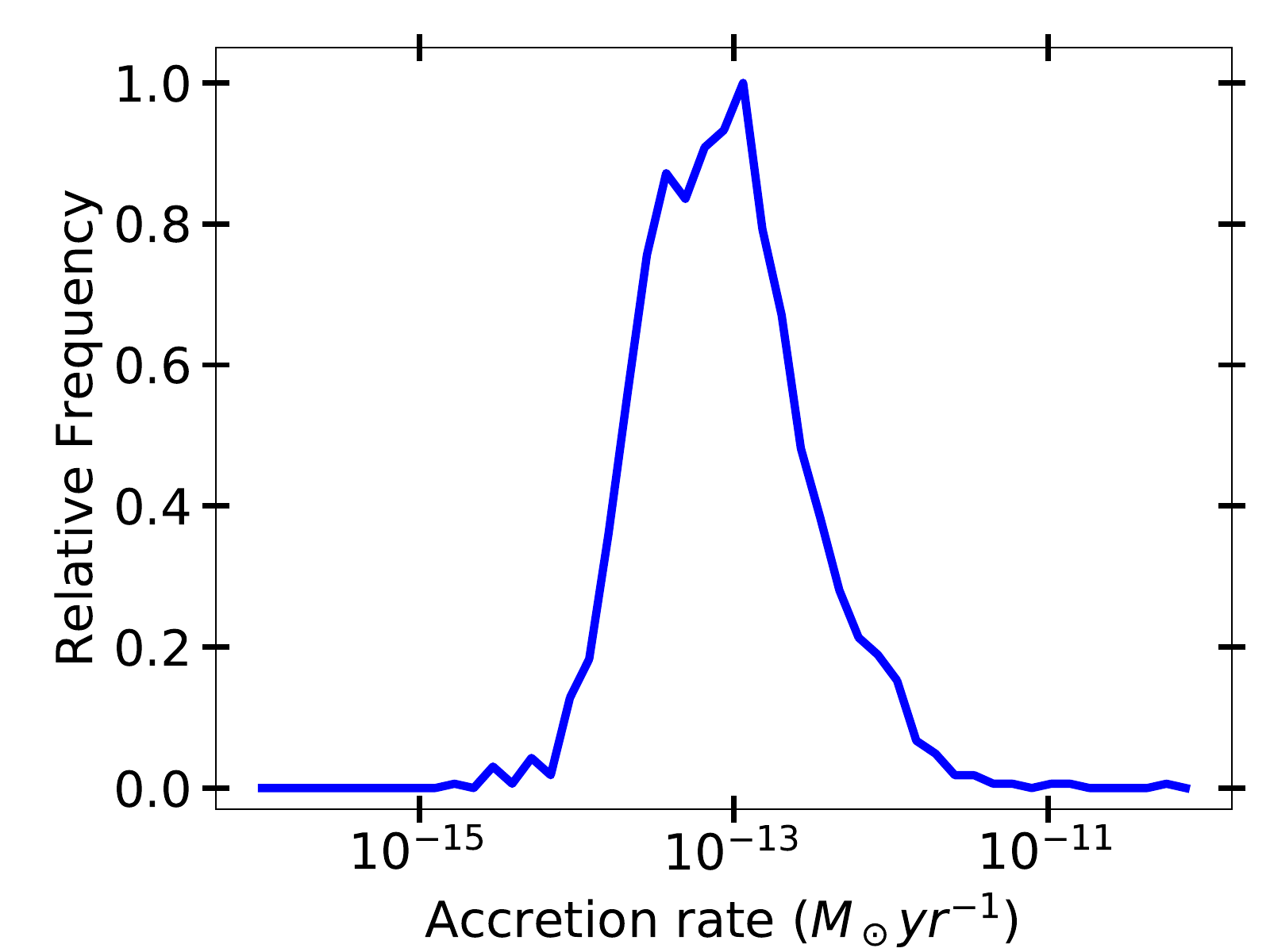}
\caption{The accretion rates for stars passing through a GMCs on a Sun-like orbit.}
\label{fig:accs}
\end{figure}

As a star passes through a GMC, gas is accreted through Bondi-Hoyle-Lyttleton accretion. The accretion rate is given in the equation below\citep{Hoyle1939, Bondi1944}.

\begin{equation}
\dot{M} = \frac{4\pi G^2M_*^2\rho}{v_{\rm rel}^3}
\label{eq:bh}
\end{equation}
Where $M_*$ is the mass of the star that passes through a GMC which has density $\rho$  with velocity $v_{\rm rel}$. This simple analytic approach has been studied in simulations by e.g. \cite{Springel2005, Rafferty2006} and they show that the accretion rates drop below the analytic one once feedback is considered. However, the temperatures and time-scales for which this becomes relevant are much higher and longer than what we'd expect to see for a GMC hit, therefore we can take the hit history of the Sun-like trajectories and apply equation \ref{eq:bh} to them in order to determine the typical accretion rates experienced by the Sun when it passes through a GMC. The density of the GMCs is set to be the mean, i.e. the mass divided by the spherical volume. The distribution of densities is shown in figure~\ref{fig:rhos}. GMCs have a $R^{-1}\--R^{-2}$\citep{Cambresy1999, Faesi2016} density profile and the impact parameters from the hit history follow an inverse distribution to that so using the mean gives a reasonable estimate of the average accretion rates. The resulting rates can be seen in figure~\ref{fig:accs}.

From figure~\ref{fig:accs} we can conclude that during a regular GMC passage the particle densities required ($n>1000~{\rm cm}^{-3}$) to trigger so called ``Snowball Earth" as described by \cite{Pavlov2005, Yeghikyan2006, Kataoka2013} will not be reached. This is because we know from studies of protoplanetary disks that they have depletion rates exceeding $10^{-10}{\rm M}_\odot {\rm yr}^{-1}$\citep{Clarke2001}. The Sun would likely evaporate the incoming gas at a 6-8 times lower rate than a planet-forming star due to the lower UV flux \citep{Claire2012}, however the evaporation would still be large enough to prevent any buildup of gas. 

It should be pointed out that GMCs are not smooth structures \citep{Draine2011}. They contain local over-densities known as clumps the densities of these objects can be between $10^2-10^4$ times higher than the average GMC density. The accretion rates scales linearly with the surrounding gas density which means that a passage through such a clump would then cause a corresponding increase in accretion by a factor of $10^2-10^4$. We can estimate how often a star passes through a clump as it passes through a GMC by looking at the problem geometrically. Observations with APEX presented in \cite{Urquhart2018} show that a $\sim 10^6M_\odot$ GMC will host $\sim10$ clumps, and the size scale of these clumps is $\sim1$ pc. The median size of GMCs hit in our simulation is 50 pc. Looking at surface area covered we get 0.4\% probability of going through a clump. We can easily show that the accretion rates in a clump would be sufficient, if we take a typical Bondi-Hoyle radius from our simulation, $\sim 7$ AU and say that we need to fill a sphere with that radius such that the mean density exceeds $1000~{\rm cm}^{-3}$. To do that it needs to accrete $10^{-11}M_\odot$ during a clump passage, which is easily done when the photo evaporation is overcome. However, the Sun experiences $1.6\pm1.3$ GMC crossings per Gyr. For that reason it unlikely that any of the ice ages in the past 500 Myrs were caused by passing through a clump. 


\subsection{Nearby and distant supernovae}
\begin{figure}
\centering
\includegraphics[width = 1\columnwidth]{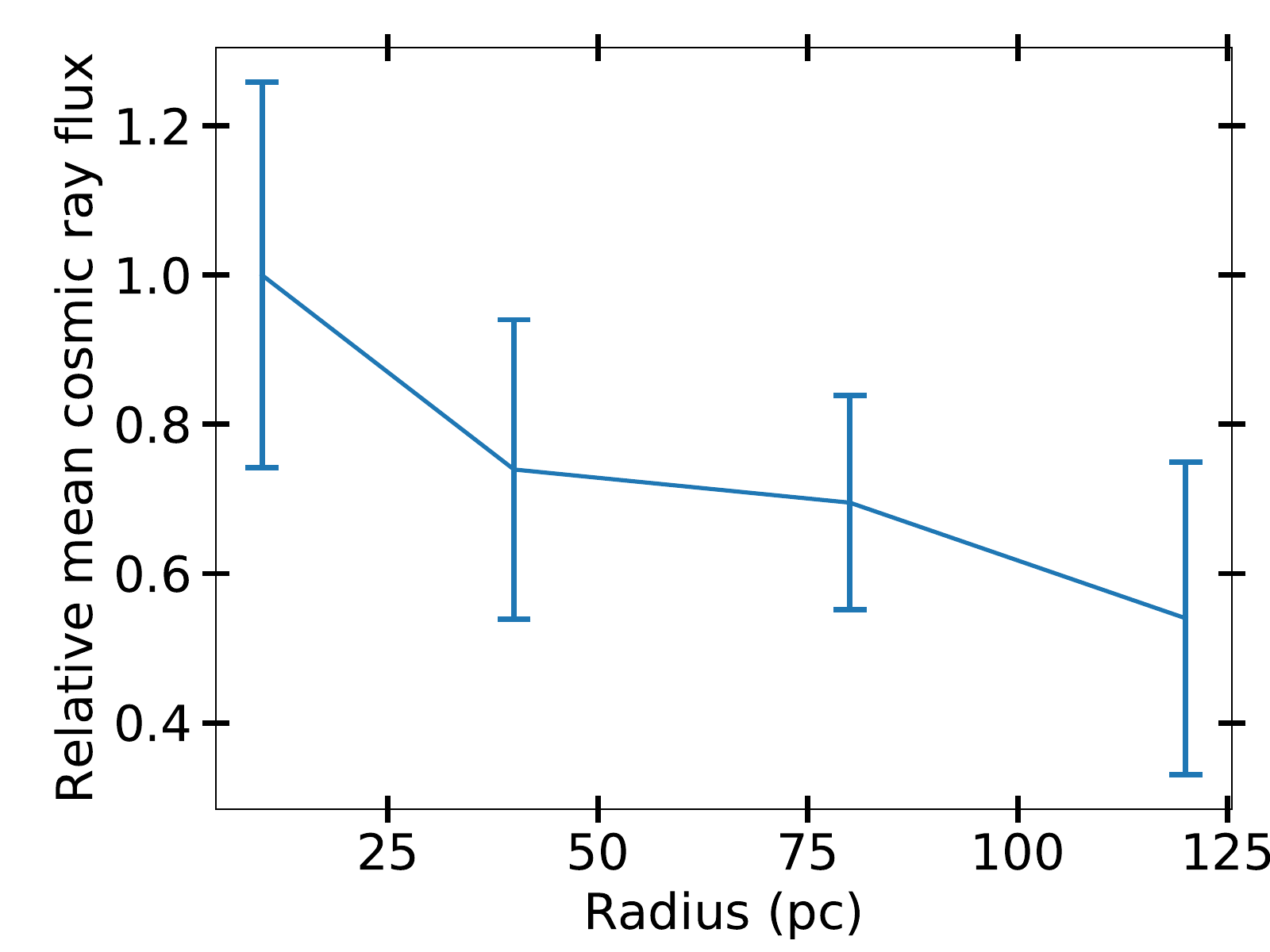}
\caption{The total cosmic ray flux from supernovae at different distances from a star orbiting through the Galaxy, calculated using equation~\ref{eq:crf} .}
\label{fig:relsne}
\end{figure}

We have looked at the the occurrence rate of supernovae within 10 pc as it has been shown that this is the distance at which a mass extinction can be caused by a single supernovae \citep{Beech2011}. Other work \citep{Shaviv2002, Shaviv2014} has pointed out that the integrated flux of cosmic rays from the more distant supernovae can also be damaging to the habitability of a planet. For this reason we compare the two effects by looking at the total cosmic ray flux as a function of distance from a star. We do this by modifying the procedure first described in section~\ref{sec:supernova}. We do the calculation using the median GMC mass encountered by the Sun ($5\times10^5M_\odot$) with the median encounter velocity for the Sun (14 km/s). However, instead of drawing trajectories just through the GMC we let them pass through or near it, where the maximum distance they can pass from the surface of the GMC is also the maximum distance at which we count the supernovae. We determine the number of supernova occurring within 10 pc thick spherical shells. To get an estimate of the flux we use the equation below.
\begin{equation}
F(R) = \Gamma(R)\sum_{i=0}^Nd_i^{-2}
\label{eq:crf}
\end{equation} 
\noindent Here $d_i$ is the distance to a supernova and $\Gamma$ is the relative frequency of the GMC encounters as a function of radius as determined in section \ref{sec:gr}. The result is shown in figure~\ref{fig:relsne}. In it we see that over time the flux from the nearby supernovae will be the dominant source of cosmic rays hitting the star. The difference is not that large, it is a factor of $\sim2$ between the innermost and outermost radii. In the outer parts of the Galaxy several Gyrs can pass without a supernova occurring within 10 pc so to fully understand the habitability of the Galaxy both the nearby and distant supernova need to be considered.

\subsection{Non-axisymmetric potentials}\label{sec:pots}
\begin{figure}
\centering
\includegraphics[width = 1\columnwidth]{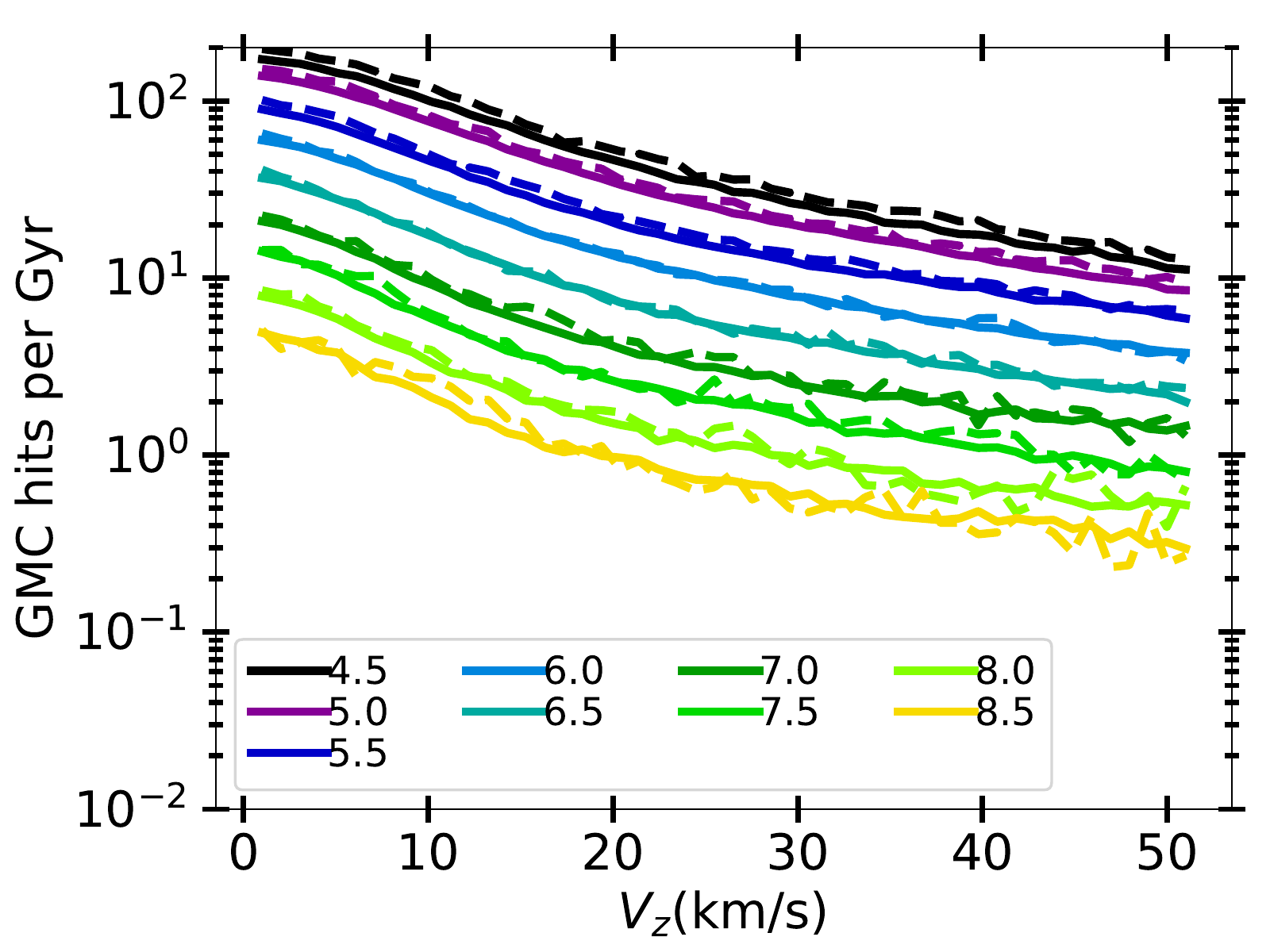}
\caption{The number of GMC hits per Gyr for runs with a spiral arm potential (dashed) and without (solid) as a function initial $V_z$. The stars were initialized at different radii using spiral arm model A and then binned and plotted in 500 pc wide bins centred at the values shown in the figure. Velocity dispersion is $\sigma_R$ = 30 km/s and $\sigma_\theta$ = 20 km/s. It should be noted that for computational efficiency the number of trajectories in the runs with the spiral potential are reduced by a factor of ten. This lead to large amounts of noise at large $V_z$ for $R>6$ kpc and bins with $R>8.5$ kpc have been excluded as they were too noisy.}
\label{fig:wa}
\end{figure}

We estimate the effect a spiral arm potential would have had by implementing a modified version of the potential presented in \cite{Pichardo2003}. In their work, they construct the spiral arm potential by super-positioning oblate spheroids along the arm. We take their model and construct an analytical version of it in which the oblate spheroids are replaced by Plummer spheres (equation~\ref{eq:plum}). We put in 50 Plummer spheres per arm, and set the total mass of the arms consisting of 200 Plummer spheres arms to be 1.75\% of the disk mass shown in table~\ref{tab:Potparam}. We set the softening length ($b_{b,h}$ in equation~\ref{eq:plum}) to be 500 pc, the same as the semi-minor axis of the oblate spheroids in \cite{Pichardo2003} and with the mass of the Plummer spheres following the same radial distribution as the gas. We find that the effect on the number of GMC hits is small, as the gravitational focusing by the arms is limited. Figure~\ref{fig:wa} shows an increase in the number of GMC hits by 5-10\% at a given radius which is small compared to the spread in the number of hits seen in figure~\ref{fig:hithist}. The only significant difference that we do see is that there are fewer trajectories in the bins at large radii. The spiral arms scatter the trajectories such that their guiding radii on average become slightly smaller. This effect is \textit{not} churning, or as it is also known, radial migration. For that co-rotation with the spiral arms is needed, along with transient spiral arms \citep{Sellwood2002}. Which is not the case here since with our potential and chosen pattern speed ($20 {\rm km~s}^{-1}{\rm kpc}^{-1}$) co-rotation occurs at 11.6 kpc. The inclusion of the spiral arms also leads to a marginally higher velocity dispersion which slightly reduces the supernova risk during a GMC hit as seen in figure~\ref{fig:snevel}, this effect is however also quite small. All in all the inclusion of the spiral arm potential would have a very small effect on the result.

Looking specifically at radial migration. Studies {\citep{schb, Frankel2018} have suggest that the Sun formed at 5 kpc and then migrated outward. From the previously determined radial scale length of the number of hits per Gyr of 750-850 pc we find that this change corresponds to 3.5-4 scale lengths depending on the $V_z$. This means that such a migration would result in a factor of 35-55 change in the hit rate depending on the $V_z$. As was discussed with respect to figure~\ref{fig:hithist}, at 5 kpc all trajectories of young stars (low $V_z$) have several tens of to hundreds hits per Gyr. This means that prior to it's migration the Sun was constantly hitting GMCs and being affected by supernovae, whereas at 8 kpc it's not uncommon for a Gyr to pass without a single GMC being hit. This does not take into account that the SFR was higher by a factor of 3 \citep{vanDokkum2013} in the Sun's infancy (4.5 Gyrs ago). The Sun's birth at 5 kpc and subsequent migration could in part explain the late emergence of land-based life on the Earth.

As for the potential itself, as stated previously we tested a set of potentials \citep{Paczynski1990, Millan2017} and only found differences in the hit rate on the order of 10\% when reproducing the simulations in different potentials. The total number of hits remained largely unchanged however we do see some local differences, i.e. there is a variation in the number of hits  at a given $V_z$ for different radii. This can be explained by the fact that the motion of the GMCs and stars change in the same way when the potential is changed. What is invariant between the potential changes is the distribution of the GMCs, e.g. in one potential at a given $R$ and $V_z$ a star might get out of the GMC layer whereas in a different one it just barely cannot. This will of course lead to some differences. 

\subsection{Galactic Habitable Zone}\label{sec:ghz}
The Galactic habitable zone (GHZ) is an annular region in the Galaxy in which one is most likely to find a habitable planet~\citep{Gonzalez2001}. The outer edge of the annulus is generally determined by the average metallicity of stars, once this drops too low ([Fe/H]$\lesssim -0.7$) making an Earth-analogue which can sustain complex life becomes impossible \citep{Lineweaver2004}. The inner edge of the annulus can also be set by looking at the metallicity. A high metallicity results in a higher occurrence rate of gas giants \citep{Fischer2005}, this can inhibit the chances of forming terrestrial planets in the habitable zone \citep{Levison2003} and if planets do end up in the habitable zone their orbits can easily be destabilized \citep{Carrera2016, Agnew2017, Agnew2018}. Additionally, authors argue \citep{Gonzalez2001, Vuktoic2018} that the high metallicity will make terrestrial planets with such high surface densities as to be inhospitable for organic life. Another way of setting the bound is by looking at the hazards posed by the Galaxy, primarily nearby supernovae \citep{Gowanlock2011, Spitoni2014, Morrison2015}.

Using our results we can determine our own habitable zone in which the z-dimension (which is often excluded) is included in the analysis and supernovae are correctly distributed in the Galaxy. The Sun is obviously habitable, so we take our findings for Sun-like orbits and determine a habitability-criterion based on that. The Sun experiences $1.6\pm1.3$ GMC hits per Gyr so as the habitability criteria we say that a star must experience two or fewer GMC hits per Gyr. Then we determine the distribution of the number of habitable stars as a function of Galactic radius using the equation below:

\begin{equation}
N = \sum_{V_z} A\times f_{\rm 2}(V_z, R)\times f_{\rm h}(V_z)\times f_{\rm [Fe/H]}(R) \times\Sigma(R) Rdr
\end{equation}
\label{eq:hab}
Where A is a normalization constant, $f_{\rm 2}$ the fraction of runs at a given $V_z$ and R that have fewer than two hits per Gyr. $f_{\rm h}$ is the fraction of stars in a $V_z$-bin given a scale height of the stars. $\Sigma(R) Rdr$ is the number of stars in a radial bin, scale height and length given by \cite{Bland-hawthorn2016}. $f_{\rm [Fe/H]}$ is the fraction of stars at $[Fe/H]>-0.7$, given by \cite{Hayden2015}. The resulting distributions can be found in figure~\ref{fig:hab}. For the thin disk we find the habitable zone to be at 5.8 to 8.8 kpc with a peak at 7 kpc, and for the thick disk 4.5 to 7.7 with a peak at 5.7 kpc, with a peak at 17\% relative to the thin disk. The thick disk is the kinematic thick disk with a radial scale length of 2 kpc and a scale height of 1 kpc. The width of what we call the habitable zone is where the occurrence rate drops to 50\% of the peak value. 

Further, one could consider where one is most likely to find a star having a planet hosting complex life on it. For this one would also have to consider the age of the systems as it takes time for life to evolve which means that the longer time a system has been within the GHZ the higher the probability of of complex life having formed. From \cite{Martig2016} we know there is a strong negative radial age gradient, i.e. there are more older stars at small radii. This means that both the distributions in figure~\ref{fig:hab} would shift inwards and additionally the thick disk can even host more habitable planets depending on ones criteria as it has a significantly older population.
\begin{figure}
\centering
\includegraphics[width = 1\columnwidth]{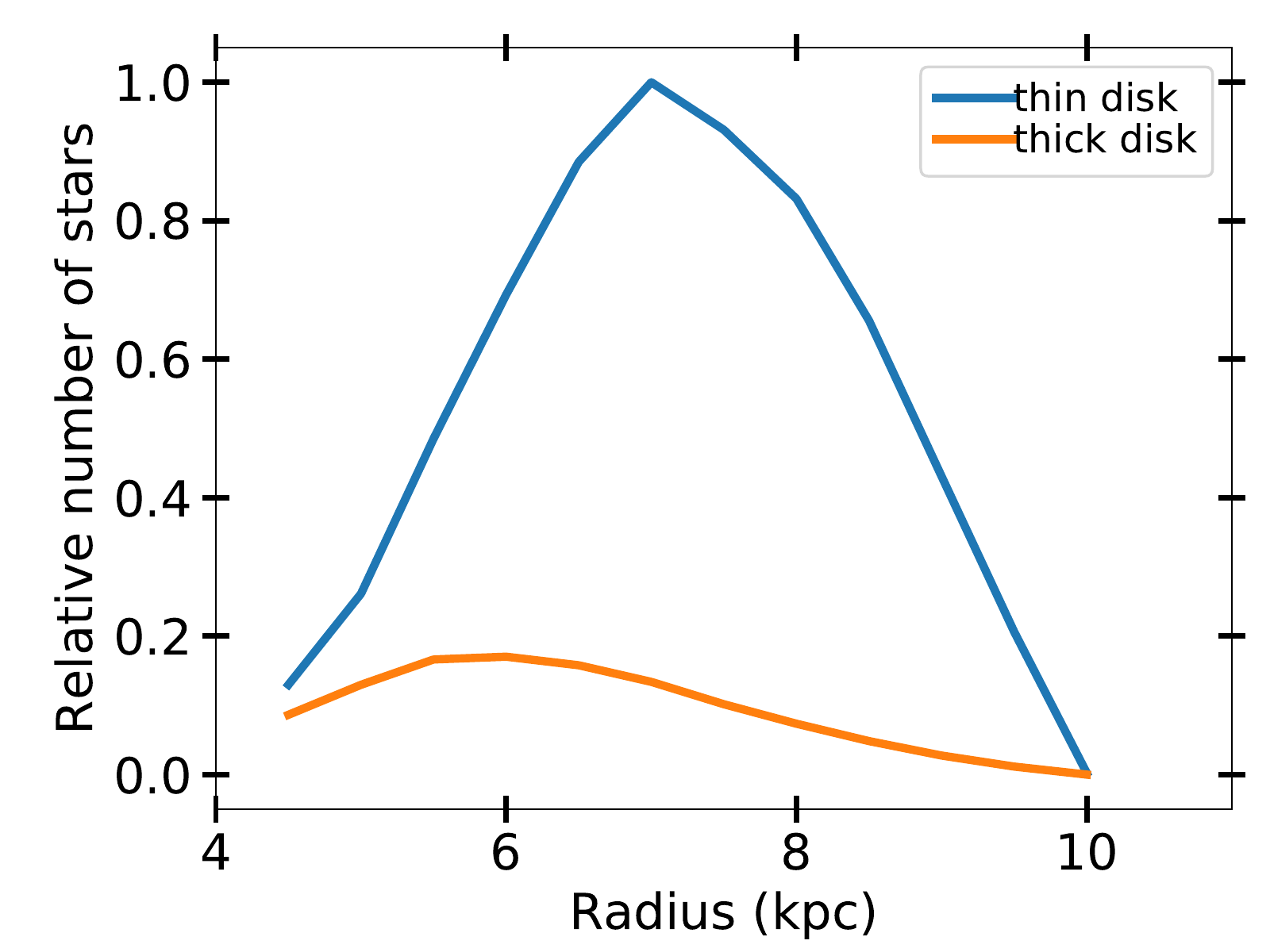}
\caption{The relative number of habitable stars in the Galaxy as a function of radius. This is calculated using the condition for habitability that a star hits fewer than two GMCs per Gyr, which is based on the Solar value. Then, using the structure of the disks found in \protect\cite{Bland-hawthorn2016} and equation \protect\ref{eq:hab} we determine the distributions. }
\label{fig:hab}
\end{figure}
\section{Summary}
We have investigated how often stars on different trajectories pass through GMCs (hits) and during a passage how often a star passes within 10 pc of a supernova as it explodes and is affected by it. Through this we have been able to address a number of issues both with regards to the Sun and its history but also with regard to the Galaxy and other stars as a whole. The main results are:
\begin{enumerate}[I]

\item We find that the number of GMC hits decrease exponentially with increasing radius ($R$) and velocity perpendicular to the Galactic plane ($V_z$). The resulting hit rates can be seen in figure~\ref{fig:hithist}. We find the scale length for the number of hits to lie between 750 and 850 pc dependent on the $V_z$ and the scale velocity is $\sim12$ km/s (table \ref{tab:fits}). These effects can cancel out, meaning that a star at 7 kpc with a $V_z$ of 5 km/s will experience the same number of hits as a star at 5 kpc with a $V_z$ of $\sim30$ km/s. \\

\item We find a narrow distribution in the number of GMC hits at smaller radii for low $V_z$ orbits. This implies that at 4-5 kpc nearly all stars will experience several tens of GMC hits per Gyr whereas a significant fraction of stars at 8 kpc experience no GMC hits at all during one Gyr, even though the mean is roughly two. \\

\item As a star ages and $V_z$ is excited the frequency of GMC hits will go down and the frequency of supernovae a star is affected by will go down even more rapidly (figure~\ref{fig:SNfrac}) because when the star does hit a GMC it will pass through it with a higher velocity and thus decrease the chance of being affected by a supernova because the risk is dependent on the velocity with which the GMC is hit.\\

\item We find that a Sun-like orbit hits $1.6\pm1.3$ GMCs per Gyr (once every $625^{+2700}_{-280}$ Myrs) and we find that it is within 10 pc of and affected by a supernova $1.5\pm1.1$ times per Gyr (once every $667^{+2500}_{-240}$ Myrs) or $0.8\pm 0.6$ times per Gyr if one corrects for the spatial and temporal clustering, i.e. if a star passes through a GMC as supernovae are going off it is likely to be hit by more than one (once every $1.25^{+3.75}_{-0.55}$ Gyrs). \\

\item The nearby supernova occurrence rate that we calculate differs somewhat from literature values. This is due to the fact that previous authors have used the local stellar density to determine the nearby supernova occurrence rate, we argue that it is the ${\rm H}_2$ density as a proxy for star formation rate that matters and in figure~\ref{fig:expR} we see the resulting differences in using a stellar and gas density for the ocurrence rates.\\

\item If the Sun was located at 5 kpc it would have a much higher GMC hit frequency and therefore also similarly be affected much more by supernovae. It has been suggested that the Sun did form at 5 kpc \citep{schb, Frankel2018}, if this is the case it will constantly have been perturbed by the Galaxy at earlier times. \textit{A migration later in the lifetime of the Sun could explain the late emergence of land-based life on Earth.}\\

\item We find that our model is consistent with deposits from recent distant supernova history traced by radioactive elements found on ocean floor. We show that the observed local arm substructure explains the supernova record well and we favour a model in which both the recorded supernovae originated from the same star forming cloud. To model the history of the Sun specifically one needs to include the potential of the spiral arms.\\

\item The amount of gas needed to be accreted by the Sun to trigger a ``Snowball Earth" is easily reached based on our calculations. However, the main obstacle is overcoming the evaporation of incoming gas by the Sun. To do so a much larger accretion rate is required, which can be achieved in very few cases (0.4\%) when a clump will be hit resulting in accretion rates up to $10^2-10^4$ times the normal.\\

\item Applying our data and results to \cite{Napier1982, Hut1984} we find that the Oort Cloud has been depleted to only contain $10^{-2}-10^{-3}$ of its original comets. However, we find that the depletion is very sensitive to the number of and the relative velocity of the encounters; implying that some Sun-like stars can have nearly pristine Oort Clouds and others very depleted. \\

\item We determine the habitable zone for Galaxy for the two kinematically different disks. We do this by using the hit history for the Sun-like orbits (figure~\ref{fig:sunhits}) to find limits for habitability, along with average metallicity of the Galaxy and stellar number-count. For the thin disk we find it to lie between 5.8-8.8 kpc and for the thick disk between 4.5-7.7 kpc peaking at 7 kpc and 5.7 kpc respectively as can be seen in figure~\ref{fig:hab}. 

\end{enumerate}

\section*{Acknowledgments}
The authors are supported by the project grant 2014.0017 ``IMPACT" from the Knut and Alice Wallenberg Foundation. The simulations were performed on resources provided by the Swedish National Infrastructure for Computing (SNIC) at Lunarc, which we can contribute thanks to grants from The Royal Physiographic Society of Lund. We would also like to thank the referee who gave us valuable input that helped improve the paper.

\bibliography{MISS.bib}

\bibliographystyle{mnras}

\end{document}